\pdfoutput=1

\documentclass[10pt]{article}

\usepackage{listings}    
\usepackage{graphicx}    
\usepackage{subcaption}  
\usepackage{algorithm}   
\usepackage{algorithmic} 
\usepackage{booktabs}    
\usepackage{hyperref}    
\usepackage[htt]{hyphenat} 
\usepackage{microtype} 
\DisableLigatures{}    

\usepackage{caption}
\usepackage{setspace}
\usepackage{multirow}
\usepackage{enumerate}
\usepackage{pdflscape}
\usepackage{pgfplots} 

\usepackage{xcolor}
\definecolor{codegreen}{rgb}{0.25,0.5,0.35}
\definecolor{codegray}{rgb}{0.5,0.5,0.5}
\definecolor{codepurple}{rgb}{0.6,0,0}
\definecolor{backcolour}{rgb}{0.95,0.95,0.92}
\definecolor{colorstring}{rgb}{0.5,0,0.35}
\definecolor{rltred}{rgb}{0.5,0,0}
\definecolor{rltgreen}{rgb}{0,0.5,0}
\definecolor{rltblue}{rgb}{0,0,0.5}
\definecolor{DarkGreen}{rgb}{0.00,0.60,0.00}
\definecolor{ScarletRed}{rgb}{0.80,0.00,0.00}
\definecolor{blizzardblue}{rgb}{0.67, 0.9, 0.93}
\definecolor{green-yellow}{rgb}{0.68, 1.0, 0.18}
\definecolor{dkgreen}{rgb}{0,0.6,0}
\definecolor{gray}{rgb}{0.5,0.5,0.5}
\definecolor{mauve}{rgb}{0.58,0,0.82}
\definecolor{lightgrey}{rgb}{0.90,0.90,0.90}
\definecolor{grey}{gray}{0.75}
\definecolor{light-gray}{gray}{0.80}

\lstdefinestyle{mystyle}{
    escapechar=©, 
	backgroundcolor=\color{backcolour},
    basicstyle=\footnotesize\ttfamily,
   	identifierstyle=\footnotesize\ttfamily,
	commentstyle=\color{codegreen},
	keywordstyle=\color{colorstring}\bfseries,
	numberstyle=\ttfamily\color{codegray},
	stringstyle=\ttfamily\color{DarkGreen},
	breakatwhitespace=false,
	breaklines=true,
	captionpos=b,
	keepspaces=true,
	numbers=left, 
	numbersep=2pt,
	showspaces=false,
	showstringspaces=false,
	showtabs=false,
	tabsize=2
}
\usepackage{xspace}
\newcommand{\evo}{{\sc EvoMaster}\xspace}

\usepackage{boxedminipage}
\newenvironment{result}%
{\smallskip
	\noindent
	\let\emph=\textbf
	\begin{boxedminipage}{\columnwidth}\begin{center}\em}%
		{\end{center}\end{boxedminipage}%
}

\usepackage{ifthen}
\newboolean{showcomments}
\setboolean{showcomments}{true} 

\ifthenelse{\boolean{showcomments}}{
	\newcommand{\nbc}[3]{
		{\colorbox{#3}{\bfseries\sffamily\scriptsize\textcolor{white}{#1}}}
		{\textcolor{#3}{\sf\small$\langle$\textit{#2}$\rangle$}}}
	
}{
	\newcommand{\nbc}[3]{}

}

\newcommand{\catwatch}{{\emph{catwatch}}\xspace}

\newcommand{\ocvnrest}{{\emph{ocvn-rest}}\xspace}

\usepackage{authblk}                
\usepackage[square,numbers]{natbib} 
\usepackage{amsmath}                

\usepackage{geometry}
\title{
Advanced White-Box Heuristics for Search-Based Fuzzing of REST APIs
}

\author[1,3]{Andrea Arcuri}
\author[1]{Man Zhang}
\author[2]{Juan Pablo Galeotti}

\affil[1]{Kristiania University College, Norway}
\affil[2]{University of Buenos Aires and CONICET, Buenos Aires, Argentina}
\affil[3]{Oslo Metropolitan University, Norway}

\date{}

\begin{document}

\maketitle

\begin{abstract}
Due to its importance and widespread use in industry, automated testing of REST APIs has attracted major interest from the research community in the last few years.
However, most of the work in the literature has been focused on black-box fuzzing.
Although existing fuzzers have been used to automatically find many faults in existing APIs, there are still several open research challenges that hinder the achievement of better results (e.g., in terms of code coverage and fault finding).
For example, under-specified schemas are a major issue for black-box fuzzers.
Currently, \evo is the only existing tool that supports white-box fuzzing of REST APIs.
In this paper, we provide a series of novel white-box heuristics, including for example how to deal with under-specified constrains in API schemas, as well as under-specified schemas in SQL databases.
Our novel techniques are implemented as an extension to our open-source, search-based fuzzer \evo.
An empirical study on 14 APIs from the EMB corpus, plus one industrial API, shows clear improvements of the results in some of these APIs.

\end{abstract}


\section{Introduction}

RESTful APIs are widely used in industry to build services that are available on the internet.
At the time of this writing, there are several thousands of REST APIs that are either free or commercial~\cite{ProgrammableWeb,APIsguru,RapidAPI}.
Besides providing functionality over the internet, this kind of APIs are also very common in industry for building enterprise systems with \emph{microservice architectures}~\cite{newman2021building,rajesh2016spring}.

Testing this kind of systems is challenging, due to communications over a network (typically HTTP over TCP), plus all issues in dealing with their environment (e.g., SQL databases and network communications with other APIs).
Writing system tests for RESTful APIs can be a tedious, time-consuming, error-prone  activity if done manually.
In industry, there is a concrete need to automate this task~\cite{arcuri2018experience}.

Due to these challenges and practical importance in industry, in recent years there has been a significant amount of research on this topic~\cite{golmohammadi2023testing}.
Several \emph{black-box} fuzzers have been implemented and evaluated, such as for example (in alphabetic order):
bBoxrt~\cite{laranjeiro2021black},
\evo~\cite{arcuri2018evomaster}
Restest~\cite{martinLopez2021Restest},
RestCT~\cite{wu2022icse},
Restler~\cite{restlerICSE2019},
RestTestGen~\cite{viglianisi2020resttestgen}
and
Schemathesis~\cite{hatfield2022deriving}.
This kind of black-box fuzzers have been used to find several real faults in existing APIs.
However, there are still many \emph{open problems} in this testing domain~\cite{zhang2023open}, which significantly hinders the achievement of better results (e.g., code coverage and fault finding).

One major issue is that often the \emph{schemas} of these APIs are \emph{under-specified}~\cite{marculescu2022faults,zhang2023open}.
For example, if one query parameter in a REST endpoint is not specified in the schema, there is no (current) way for a black-box fuzzer to generate test cases using such a query parameter.
Creating query parameters with random names would have an extremely low probability of matching any of these existing unspecified parameters.
This can reduce the chances of achieving higher code coverage if such parameters have a major impact on the execution control flow of the API.

At the time of this writing, our \evo fuzzer is the only tool that supports \emph{white-box} testing of RESTful APIs (besides as well RPC~\cite{zhang2023rpc} and GraphQL~\cite{belhadi2023random} APIs).
In our previous empirical studies~\cite{zhang2023open}, as well as in independent studies~\cite{Kim2022Rest} from other researchers, it has been shown that white-box testing for REST APIs achieves better results than black-box fuzzing.
Furthermore, even in black-box mode~\cite{arcuri2020blackbox}, \evo achieved the best results (with Schemathesis~\cite{hatfield2022deriving} having similar results) in these tool comparisons~\cite{zhang2023open,Kim2022Rest}.
Considering that \evo has been successfully used on large industrial, microservice enterprise systems with millions of lines of code~\cite{zhang2023rpc}, finding thousands of real faults that have been confirmed to be fixed by their developers throughout the years, it can be arguably considered the state-of-the-art in this testing domain.
Still, there are several research challenges to address, as large parts of the code-base of these APIs are not covered with existing techniques~\cite{zhang2023open,zhang2023rpc}.

To address these research challenges, in this paper we provide a series of novel \emph{white-box} heuristics to improve the performance of fuzzing in this testing domain.
In particular, we aim at addressing three different problems:
\begin{enumerate}
\item \emph{flag} problem~\cite{BaS03} in common library calls~\cite{arcuri2021enhancing};
\item under-specified schemas in OpenAPI definitions, in particular when dealing with missing HTTP query parameters and headers information;
\item under-specified constraints in SQL database schemas, in particular when the tested API uses the Java Persistence API (JPA) to access the databases.
\end{enumerate}

Our novel white-box techniques can be used and applied to any white-box fuzzer.
For the analyses in this paper, we have implemented these techniques as part of our \evo fuzzer, in particular focusing on the JVM (e.g., for APIs written in either Java or Kotlin).
However, they could be adapted to other programming languages as well, such as for the white-box fuzzing of JavaScript/TypeScript APIs running on NodeJS~\cite{zhang2023javascript} and C\# APIs running on~.NET~\cite{golmohammadi2023net}.

To validate our novel techniques, in our empirical study in this paper we used all the 14 open-source, JVM REST APIs currently part of the EMB corpus~\cite{icst2023emb}.
To better generalize our results, in our empirical study we also included 1 closed-source, industrial API from one of our industrial partners.
Our empirical study shows that, in some of these APIs statistically significant improvements were achieved (both in terms of code coverage and fault detection).
On the one hand, this enables us to push forward the boundaries of scientific research in white-box fuzzing of Web APIs.
On the other hand, there are still several challenges left, which will require further research to design more advanced white-box heuristics to solve these further issues.

The paper is organized as follows.
Section~\ref{sec:background} provides background information to better understand the rest of paper.
Related work is discussed in Section~\ref{sec:relatedwork}.
Our new handling of the flag problem is discussed in Section~\ref{sec:new_replacements},
followed by how we deal with under-specified schemas for OpenAPI in Section~\ref{sec:schemas},
and for SQL in Section~\ref{sec:sql}.
Section~\ref{sec:timed_events} provides our solution to the issue given by timed events.
Our empirical study is presented in Section~\ref{sec:study}.
Discussions on the obtained results follow in Section~\ref{sec:discussion}.
Threats to validity of our study are analyzed in Section~\ref{sec:threats}.
Finally, Section~\ref{sec:conclusions} concludes the paper.

\section{Background}
\label{sec:background}

\subsection{REST APIs}
\label{sub:}

Presently, a significant portion of web services are implemented following the REST (REpresentational State Transfer) architectural style~\cite{fielding2000architectural}. 
Notable adopters of this approach among organizations and companies are Google\footnote{https://developers.google.com/drive/v2/reference/}, Amazon\footnote{http://docs.aws.amazon.com/AmazonS3/latest/API/Welcome.html}, Twitter\footnote{https://dev.twitter.com/rest/public}, Reddit\footnote{https://www.reddit.com/dev/api/}, LinkedIn\footnote{https://developer.linkedin.com/docs/rest-api}, among others. 
Beyond their role in furnishing internet-based functionalities (e.g., see API portals such as
\emph{APIs.guru}\footnote{https://apis.guru/} and \emph{RapidAPI Hub}\footnote{https://rapidapi.com/hub}), the REST architectural style also holds sway within enterprise backends, particularly when microservices architecture are used~\cite{newman2021building,rajesh2016spring}.

The REST architectural style is not a protocol per se, but rather a collection of principles governing and structuring  resources accessible over HTTP(S) networks. 
Resources are pinpointed via URLs and can be manipulated using HTTP semantics, with \texttt{GET} requests for data retrieval, \texttt{POST} for data creation, \texttt{PUT/PATCH} for data modification, and \texttt{DELETE} for deletion. 
Inputs can be transmitted through path components in URLs, query parameters, HTTP headers, and payload bodies. Various data formats can be employed for transmission, with JSON (JavaScript Object Notation) currently ranking among the most prevalent such data format.

To enhance the understability and usability of REST APIs, a common strategy involves providing \emph{schemas} outlining the available endpoints and supported input formats (including details like query parameter names and types). Multiple schema standards exist, with OpenAPI/Swagger~\cite{Swagger} standing out as the current industry leader. This format harnesses either JSON or YAML (Yet Another Markup Language) to specify these schemas. Schemas can be authored either manually or auto-generated from API source code (exemplified by tools like SpringFox and SpringDoc for the widely adopted enterprise framework Spring~\cite{Spring}).

\subsection{EvoMaster}
\label{sub:}

\begin{figure}[!t]
\begin{lstlisting}[language=java,basicstyle=\footnotesize]
@Test
public void test() throws Exception {
 // Prepare the request body for the POST request in JSON format
 String requestBody = "{\"firstName\": \"Alice\", \"lastName\": \"Smith\"}";

 // Create the new resource (POST request)
 String newResourceId =  given()
    .contentType(ContentType.JSON) 
    .body(requestBody)
  .when()
    .post("/users")
    .then()
      .statusCode(201) // Assert the resource was created
      .extract().path("id"); // Extract the newly created resource's ID

  // Retrieve the newly created resource (GET request)
  given()
   .when()
      .get("/users/" + newResourceId)
   .then()
     .statusCode(200) // Assert resource was found
     .assertThat()
       .body("firstName", equalTo("Alice"))  // Check the expected first name 
       .body("lastName", equalTo("Smith")); // Check the expected last name
}       
\end{lstlisting}
\caption{\label{fig:rest-assured}
A system-level test case using the RestAssured library.
}
\end{figure}

In the context of REST APIs, a system-level test case is a sequence of HTTP calls to different endpoints.
The expected result of each HTTP call can be asserted, checking that the value of the status code is within some given set of expected status codes values.   
As one might expect, writing a specific system-level test case that triggers a specific behaviour in the implementation of the API (such as creating a resource with a specific data format, or retrieving an existing resource against its expected value) is difficult and time consuming.
\autoref{fig:rest-assured} presents a system-level test case (written using the RestAssured~\cite{RestAssured} library) that creates a new \texttt{User} resource using the endpoint \texttt{/users}.
Subsequently, the test case retrieves a \texttt{User} resource by means of a \texttt{GET} HTTP call, and the retrivied data is compared against their expected values. 
The automation of creating such system-level test cases remains elusive~\cite{arcuri2018experience}. 

The open-source tool \evo~\cite{arcuri2018evomaster,arcuri2021evomaster} aims at automatically  generating system-level test cases for REST APIs. 
It implements several evolutionary algorithms (such as MIO~\cite{arcuri2018test}) to evolve test cases towards maximing code coverage and fault finding metrics. 
As API calls might depend on each other (e.g., in \autoref{fig:rest-assured} a typical \texttt{GET} template is exercised, where a unique \texttt{PUT} call is exercised followed by a \texttt{GET} call), \evo can exploit dependencies among API resources~\cite{zhang2021resource,zhang2019resource} to cover more API behaviour.

\evo is divided in two main modules: 
(1) a \emph{core} process and (2) a \emph{driver}. 
The core process contains all the basic functionality for a SBST tool, like the search algorithms, fitness functions, writting the test generation outputs, etc.
On the other hand, the driver is provided as a library, which engineers need to use to specify how to start, stop and reset the SUT.
This is done with short configuration classes, that need to be implemented manually. 
However, the driver is also responsible of instrumenting the bytecode, which is done automatically.
This instrumentation allows \evo's driver to retrieve several different SBST heuristics like the branch distance.
This heuristic is not only retrieved for predicates in the control flow of the SUT, but also for all SQL commands executed over a database (if any)~\cite{arcuri2020sql}.

The core and the driver will run as separated processes, communicating over HTTP. 
This architecture will enable to support different programming languages, as a new supported language would just require a new driver library for it (such as JavaScript~\cite{zhang2023javascript,js2022} and C\#~\cite{golmohammadi2023net}).
Additionally, this architecture also allows \evo to support both white-box~\cite{arcuri2019restful} and black-box~\cite{arcuri2020blackbox} testing.
If black-box testing is chosen, \evo's \emph{core} process can directly interact with the deployed API.
Therefore, there is no need for engineers to write any driver as \evo could be run on any type of REST API regardless of their programming language.

If white-box testing is chosen, \evo currently supports APIs that run on the JVM~\cite{arcuri2018test,arcuri2019restful,arcuri2021enhancing,zhang2023open} (e.g., written in Java or Kotlin), JavaScript/TypeScript~\cite{zhang2023javascript,js2022} and C\#~\cite{golmohammadi2023net}.
For JVM-based APIs, it can output test suites in JUnit format, using the library RestAssured~\cite{RestAssured} for making the HTTP calls toward the SUT.
The generated tests will use the 
configuration classes for handling the SUT.
This means that the generated tests are self-contained: the test suite files can start the SUT before any test is run, reset the state of the SUT before/after each test case execution (to make them independent), and stop the SUT once all tests are run.
The generated test cases can be used as well for regression testing, as can be added to the repository of the SUT, and run as part of a Continuous Integration process. 

In case of black-box testing,  the generated test suites  would still be output, e.g., in either Java or Kotlin using the RestAssured library, independently of the programming language in which the target APIs were written. 

\section{Related Work}
\label{sec:relatedwork}

\subsection{Fuzzing REST APIs}
\label{subse:fuzzing}

Fuzzing is one of the most effective approaches to detect software faults, 
and to achieve higher code coverage~\cite{godefroid2020fuzzing,manes2019art,zhu2022fuzzSuvery}.
Fuzzers operate using either a white-box or a black-box approach.
In a white-box fuzzer, internal details of the system under test, such as its source code, binaries, bytecode, or SQL communication~\cite{arcuri2020sql}, will be accessed.
This information can be used to design heuristics to improve the search to produce better test cases with better code coverage and fault detection capabilities.
White-box fuzzers have been proven to be highly effective in numerous instances~\cite{godefroid2012sage,martin2021black,arcuri2020blackbox,Kim2022Rest,zhang2023open}.

Fuzzing REST APIs has gained a major amount of attention among researchers in recent years~\cite{golmohammadi2023testing}.
There are multiple fuzzing tools available for RESTful APIs~\cite{Kim2022Rest,zhang2023open}.
Some of the aforementioned \emph{black-box} fuzzers for REST APIs include (in alphabetic order):
bBOXRT~\cite{laranjeiro2021black},
\evo~\cite{arcuri2018evomaster}
ResTest~\cite{martinLopez2021Restest},
RestCT~\cite{wu2022icse},
Restler~\cite{restlerICSE2019},
RestTestGen~\cite{viglianisi2020resttestgen}
and
Schemathesis~\cite{hatfield2022deriving}.
To the best of our knowledge, \evo~\cite{arcuri2018evomaster} is the only REST API fuzzer that can be supports both white-box and black-box testing of REST APIs.
With the recent studies conducted for comparing the state-of-the-art REST API fuzzers~\cite{Kim2022Rest,zhang2023open}, \evo in white-box mode achieved the best performance in code coverage and fault detection.

\subsection{Search-Based Software Testing}
\label{sub:sbst}

Search-based software testing (SBST)~\cite{ABHP09,harman2012search} has been shown to be an effective technique to automatically generate test cases.
In SBST, the problem of 
generating adequate test suites and fault-releaving test cases 
can be reformulated as an optimization problem.
Examples include unit testing of Java software with open-source tools like EvoSuite~\cite{fraser2011evosuite}, and testing of mobile applications with the Sapienz tool at Facebook~\cite{sapienz2018}.

When doing white-box testing, different techniques are used to define heuristics to smooth the search landscape. 
The most common in the literature of SBST is the so-called \emph{Branch Distance}~\cite{Kor90}.
Given a boolean predicate in  the code of the system under test (SUT) (e.g., \texttt{x==y+10}), the branch distance provides a heuristic value (e.g., minimize the value of \texttt{|x-(y+10)|}) to guide the search towards satisfying the boolean predicate. 

Unfortunately, a common issue is the so-called \emph{flag problem}~\cite{BaS03}, where the branch distance is not able to provide any gradient.
For example, consider a scenario where our target program is dealing with string operations returning booleans~\cite{alshraideh2006search}, like comparisons of two strings for equality.
By default, a boolean predicate like \texttt{x.equals("Hello")}, would be just a flag, returning true or false.
In such a scenario, the search algorithm has no guidance whether \texttt{x} is getting closer to become \texttt{"Hello"}. 
Inputs such as \texttt{"ello"} and \texttt{"foo"} will be considered indistinguishable.
Flags in the code might depend on string operations~\cite{alshraideh2006search}, loop assignments~\cite{baresel04,binkley2011}, nested predicates~\cite{mcminnBH09}, calls to boolean functions~\cite{wapplerWB09,LiF11,lin2020recovering} and non-integer comparisons~\cite{LiF11}.

An approach to address this issue is to transform the code of the SUT to improve the fitness function, using so-called \emph{Testability Transformations}~\cite{harman2004testability}.
A testability transformation modifies the original SUT to produce a new version of the same SUT, but more amenable to the test generation process.
This change might not necessarely preserve its original semantics, but it must \emph{``preserve test sets that are adequate with respect to some chosen test adequacy criterion''}~\cite{harman2018we}. 
As an example, if a test generation targets a single goal (e.g., a specific branch in the SUT), a non-semantics preserving transformation might slice (i.e., remove) some lines/branches to alleviate a test generator.
In contrast, semantics preserving testability transformation might not (by definition) affect the chosen test adequacy criteria.
As an example, a semantic preserving testability transformation could replace the original boolean predicate \texttt{x.equals("Hello")} with a computation of a string distance, like the \emph{edit distance} (number of insertions, deletations or changes to transform a given string into another)~\cite{alshraideh2006search}.
If such a transformation was applied, comparisons to \texttt{"ello"} and \texttt{"foo"} in \texttt{x.equals("Hello")} will no longer be considered indistinguishable by the heuristic.    

Different transformations have already been proposed~\cite{HarmanBBHHKMR08,harman2018we},  mainly to deal with (but not restricted to)  
flag conditions~\cite{BaS03,gongY12,alshraideh2006search,baresel04,binkley2011,mcminnBH09,wapplerWB09,LiF11,lin2020recovering,LiF11}. 
Testability transformations have also be proposed to generate pseudo-oracles~\cite{mcminnGECCO09} (which can be helpful to detect numerical inaccuracies and race conditions).
Furthermore, besides search-based test generation, slicing-based testability transformations can also be useful to improve Dynamic Symbolic Execution test generation~\cite{converseOK17}.

\evo performs whole test suite generation (as it is done by EvoSuite~\cite{GoA_TSE12}) at the system level.
In other words, it simultaneously targets all goals within the SUT.
In this scenario, many of the aforementioned non-semantics preserving transformations  would not be applicable as they would lead to generating test cases that do behave accordingly in the original SUT.  
For this reason, a main difference in \cite{arcuri2021enhancing} and in this work, is that all of the proposed testability transformations must preserve the semantics of the SUT.

\subsection{Method Replacement and Taint Analysis}
\label{sub:taint}

In~\cite{arcuri2021enhancing} we have presented a series of testability transformations~\cite{HHH04} to improve the fitness function of \evo, which built on top of some basic examples (e.g., for \texttt{String.equals}) from EvoSuite~\cite{fraser2011evosuite}.
In particular, in the case of JDK APIs where method calls end-up providing no gradient to the search (e.g., they return either boolean values or throw an exception on invalid inputs), for several of those APIs we automatically replace those method calls with our own custom versions, at class-loading time via bytecode instrumentation.
These custom method versions are semantically equivalent: for the same inputs they give the same outputs.
However, they compute heuristic values to determine how far the test data was from returning either a \texttt{true} or \texttt{false} output, and how far they were from not throwing an exception.
Such heuristic values are then used in the fitness function to guide SBST search toward generating test data to get the desired output.

Consider this following simple example:

\begin{lstlisting}[language=java,numbers=none]
if(x.equals("A quite long string that it is unlikely to get at random")){
\end{lstlisting}

Here, sampling the string variable \texttt{x} at random would have extremely low probability of satisfying the constraint of that \texttt{if} statement.
In our bytecode instrumentation, that code would be replaced with:

\begin{lstlisting}[language=java,numbers=none]
if(StringClassReplacement.equals(x,"A quite long string that it is unlikely to get at random",targetID)){
\end{lstlisting}

Here, \texttt{String.equals} gets replaced with our own \texttt{StringClassReplacement.equals} version.
Internally, before returning the same result as \texttt{String.equals}, it computes heuristic values for the 2 possible outcome \texttt{true} or \texttt{false} (which are then registered in the search based on the label \texttt{targetID}).
Such heuristics are based on the distance values defined in~\cite{Alshraideh06}.
The search is then rewarded in the fitness function to apply modifications to \texttt{x} that lead to have at least 1 test case in which such call returns \texttt{true}, and 1 that returns \texttt{false}.

This approach gives gradient to the search to evolve a desired \texttt{x} value.
However, due to the length of the string, it can take several generations in the evolutionary process to evolve the desired value.
But what if \texttt{x} is part of the test data?
Once it is detected at runtime that such value is compared with an \texttt{equals} method, it would be more efficient to simply use the desired value (i.e., \emph{``A quite long string that it is unlikely to get at random''} in this case) directly instead of evolving it.
The problem is that \texttt{x} could be modified during the execution of the SUT (e.g., it is the result of a \texttt{substring} operation, or a concatenation of different strings).
Also, in the testing of RESTful APIs, such \texttt{x} could be the value of a URL query parameter, or a nested field in a JSON body payload in a \texttt{POST} request.

To address this issue, in~\cite{arcuri2021enhancing} we presented a technique which is a form of taint analysis.
When mutating strings in a test case (regardless of where they are, e.g., query parameters and fields in JSON objects), with a certain probability we replace such strings with tainted values in the form \texttt{\_EM\_\textbackslash{}d+\_XYZ\_}.
For example, \texttt{\_EM\_0\_XYZ\_}.
In our method replacements, every time we detect that a string input is matching such regular expression, we can directly identify which part of the genotype it comes from.
Then, the search can automatically replace such string values based on information from the replacement methods, making those constraints trivial to solve.
That regular expression is written in that way to reduce the chances that a random string in the SUT would match it.
Also most modifications on the inputs (apart from modifying digits in the middle of the string) would be detected as well.

This approach does not work in all cases (e.g., when strings are modified), but it is very effective for RESTful APIs~\cite{arcuri2021enhancing}.
The reason is that those APIs often have many string inputs that are not modified by the SUT (i.e., they are just read).

In~\cite{arcuri2021enhancing} we provided several method replacements, e.g.,
for \texttt{String.startsWith}, \texttt{Integer.parseInt} and \texttt{Collection.contains}.
For each one, we defined heuristic distances to provide gradient to the search.
However, there are several more methods in the JDK APIs that could be handled this way.

\subsection{Genotype Expansion}
\label{sub:expansion}

OpenAPI schemas could be underspecified.
For example, some query parameters and headers could be missing from the schema, albeit being handled by the API.
A white-box approach could detect some of these cases, as it can analyze the source code of the API.

In the popular Spring framework (which is the most used enterprise framework for the Java language), parameters and headers can be automatically injected via annotations in the REST controllers
(e.g., using \texttt{@RequestParam} and \texttt{@RequestHeader}).
In those cases, automated tools that create OpenAPI schemas by analyzing annotations can detect those inputs (e.g., SpringFox and SpringDoc).
However, a less common case is to pass as input to those REST controllers an object representing the whole HTTP call, such as \texttt{WebRequest}.
In those cases, automated schema generators would have no information on the actual expected structure of the incoming HTTP requests.
The generated schema definitions would hence have no useful info.

As this problem actually happens in some of the APIs in EMB~\cite{icst2023emb,EMB}, in~\cite{arcuri2021enhancing} we presented some techniques to handle some basic cases of this issue.
In particular, we provided method replacements for
\texttt{getParameter()},
\texttt{getParameterValues()},
\texttt{getHeader()}
 and
 \texttt{getHeaders()}.
Everytime any of these methods is called, and the presence of a parameter/header not in the schema is detected, our instrumentation will inform the search engine about such value.
Then, \evo will expand the genotype of the evolving individuals, creating test cases that can set and use those newly discovered parameters and headers.

Similarly to parameters and headers, there can be issues when the body payload types are unspecified.
For these cases, we provided a method replacement for  \texttt{getInputStream()} on the object representing the incoming HTTP requests.
If the execution of such call is then detected, we then trace the use of the library Gson for marshalling JSON payloads, in particular the calls to \texttt{fromJson()}.
This way, we check which classes are used as DTO to map the incoming body payloads.
This information can then be fed back to the search engine, which will now be able to evolve input objects matching those DTO structures.
As there are few technical details at play here, we refer the interested reader to~\cite{arcuri2021enhancing} for the full details on how this is achieved.

\section{New Method Replacements}
\label{sec:new_replacements}

In this paper, one of the scientific contributions is to provide method replacements with SBST heuristics for several more APIs in the JDK and common libraries.
This is a direct extension of what first presented in~\cite{arcuri2021enhancing} (recall Section~\ref{sub:taint}).
The details of those new method replacements will be discussed in the next sections.

When a new method replacement is designed, up to three different tasks need to be considered (depending on the method):

\begin{itemize}
\item   define a new \emph{branch distance}~\cite{Kor90} function $d$, to guide the search, where $d$ is minimized.
        The value $d=0$ would mean that the target is covered.
        This is done for functions that return booleans, and for functions that throw exception on invalid inputs.
        To handle some further special cases (e.g., \texttt{Map.get}) we also consider functions that might return \texttt{null}, i.e., for those cases we create two new different testing targets: returning \texttt{null} and returning a non-\texttt{null} value.
\item   if the function deals with string inputs, possibly apply taint analysis (recall Section~\ref{sub:taint}).
\item   in case of taint analysis, can define new gene types to evolve strings with specific constraints (e.g., for URL, URI and UUID, as it will be discussed in Section~\ref{sub:uri})
\end{itemize}

In \evo, each testing target (e.g., lines and branches) we try to optimize for has a heuristic value 
$h \in \left[0,1\right]$.
The value $h=1$ means that the target is covered (equivalent to $d=0$).
Note that, internally, \evo uses $h$ instead of $d$ to better handle the simultaneous optimization of many targets (e.g., with the MIO algorithm~\cite{arcuri2018test}).
Given a branch distance value $d$, $h$ is computed as 
$h= b + \left(1 - b\right)\frac{1}{1+d}$, where $b$ is a small constant, e.g., $b=0.1$ (see~\cite{arcuri2021enhancing} for full details and rationale for these choices).

Note that, for reasons of space, we cannot go here into all the low level technical details of each of these new method replacements.
Those are thousands of lines of code, to deal with many edge cases (e.g., how to deal with instances of \texttt{IdentityHashMap} which does not use \texttt{equals} for comparisons, or how to detect at runtime maps that have $\Omega(n)$ instead of $O(log n)$ complexity for \texttt{containsKey}, which would result in computational bottlenecks when calculating the branch distances).
What we present here is high level descriptions.
For full details, we refer the reader to our open-source implementation on GitHub~\cite{EvoMaster},
in particular to version 1.6.1 (which is stored on Zenodo~\cite{zenodo161evomaster} for long term storage).
In particular, most of the implementation is under the package \texttt{org.evomaster.client.java.instrumentation.coverage.methodreplacement}.

\subsection{Data Structures}
\label{sub:data-structures}

Data structures like Lists, Sets and Maps are widely used.
Many of their functions return booleans, e.g., when checking if the collection contains a specific input value, which creates fitness plateaus in the search.
In~\cite{arcuri2021enhancing}, we already handled most of those boolean methods, by providing method replacements for the class \texttt{java.util.Collection}.
One missing case was for \texttt{Collection.containsAll()}.
Given an input $Y$ for a collection $X$, such method would return \texttt{true} if every single element in $Y$ is present inside $X$.

To deal with this, we rely on the already existing distance $d_c$ defined in~\cite{arcuri2021enhancing} for \texttt{Collection.contains()}, which was defined as $d_c(e,X) = min_{x \in X}(d(x,e))$, for input element $e$ on collection $X$.
In other words, compute a branch distance on each element in the collection from the input, based on their type (e.g., string and numbers), and take the minimum (as the input just needs to match one single element, and we can prioritize the closest one).
Then, such distance $d_c$ can be converted into a heuristics value $h_c$, as previously discussed.
For the case of \texttt{Collection.containsAll()}, we then use $h_a(Y,X) = \frac{\sum_{y \in Y} h_c(y,X)}{|Y| + log |Y|}$.
Here, we compute $h_c$ for each element in $Y$ and sum them, as all of them must match one element in $X$.
As $h_c$ still must be in $[0,1]$, we scale the result of the sum by the cardinality of the collection $Y$.
However, during the search it might well be that the size of $Y$ changes (if $Y$ is an input from a test case).
The less elements in $Y$, the easier it is to satisfy such constraint.
So, an increase in length should be penalized, and that is why we add a further $log |Y|$ to the denominator.

Other missing methods are related to the removal of elements in a collection.
For example, \texttt{Collection.remove()} returns a boolean based on  whether the element was in the collection.
We apply the same distance computations as \texttt{Collection.contains()}.
Likewise, \texttt{Collection.removeAll()} is treated similarly to \texttt{Collection.containsAll()}.

Where lists and sets in the JDK do extend from \texttt{java.util.Collection}, maps do not.
In other words, the interface \texttt{java.util.Map} is not a subtype of \texttt{java.util.Collection}, although it has several methods with same name and semantics (e.g., \texttt{size()} and \texttt{isEmpty()}).
This means that none of the transformations presented in~\cite{arcuri2021enhancing} works on map data structures out of the box.
Handling this was only technical effort, as it was just a matter of implementing equivalent replacement methods for those map methods having the same semantics, such as \texttt{isEmpty()}, \texttt{get()}, \texttt{getOrDefault()}, \texttt{containsValue()}, \texttt{remove()} and \texttt{replace()}.

Although technically they are not collections, enumerations (i.e., \texttt{Enum} in Java) share some characteristics.
For example, \texttt{Enum.valueOf(s)} would return an enum instance based on the input string $s$.
If $s$ is not matching any of the values in the enumeration, it throws an exception.
Here, this is the equivalent case of checking \texttt{contains} on a list/set of strings.
So, we apply the same type of branch distance calculation (together with taint analysis, as the input is a string).

\subsection{Miscellaneous Functions}

Comparison of two objects for equality is a typical case for which heuristic distances can be defined, depending on their type, e.g., for numerical values~\cite{Kor90} and strings~\cite{AlB06}.
In Java, each object extends from \texttt{java.lang.Object}, which defines the method \texttt{equals}.
Different classes like \texttt{String} and numerics such as \texttt{Integer} and \texttt{Double} have their own overridden implementations.
We already provided method replacements for all those cases in~\cite{arcuri2021enhancing}.
However, an important missing case was when the references of the compared objects are abstracted to the root-type \texttt{Object}.
Consider the example of \texttt{x.equals(y)}, where both \texttt{x} and \texttt{y} are of type \texttt{Object}.
At compilation time, as well as at instrumentation time (when classes are first loaded into the JVM), there would be no information on their actual types.
They could be two numbers, or a TCP socket compared with a Hash Map, for what we might know.
In those cases, we would not be able to apply any of the transformations from~\cite{arcuri2021enhancing}.
However, at runtime the types of the compared objects would be known.
So, a relatively easy solution here is to provide a method replacement for \texttt{Object.equals()}.
Then, when such method is executed during the evaluation of a test case, we can check the actual, most specific types for \texttt{x} and \texttt{y} (e.g., using operators such as \texttt{instanceof}).
If those inputs happen to be of types for which we have defined any heuristic distance (e.g., strings and numbers), then we calculate such distance (as well as applying taint analysis in case the type is string).

Between what already presented in~\cite{arcuri2021enhancing}, and what presented newly in this paper, in \evo we have more than 100 methods for which we provide replacements for computing different kinds of heuristics.
However, those are applied only when such method usage can be identified in the instrumented bytecode, e.g., like a call to \texttt{x.equals(y)}.
This unfortunately leaves out the cases when methods are called by \emph{reflection},
like \texttt{m.invoke(x,y)}, where \texttt{m} is a reference to the \texttt{java.lang.reflect.Method} instance for the method \texttt{Object.equals()}.
We apply a method replacement for \texttt{Method.invoke()}, where, at runtime, we check if the used method referenced by \texttt{m} is one for which we have any replacement.
If so, inside the replacement for \texttt{m.invoke()}  we rather call the replacement for the method accessed by reflection.
Unfortunately, though, a major issue here is that reflection is used \emph{massively} in Java, especially in popular frameworks such as Spring and Hibernate.
A naive replacement for \texttt{Method.invoke()} would be (and was) a major performance bottleneck.
This was one of the few cases in which several code optimizations were needed to do not drastically reduce performance (e.g., by using different levels of caches to memoize parts of the computation).

\sloppy
Another method that can have drastic effects on performance is \texttt{sleep} in the class \texttt{java.lang.Thread}.
When a test case is evaluated, if a \texttt{sleep} is executed on the same thread of the test case, then the duration of the test case would directly increase by the amount of time given as input to \texttt{sleep}.
If the test case involves executing a network call (e.g., HTTP over TCP when fuzzing web services), such connection could timeout if the \texttt{sleep} is too long.
This is a major issue when the input of the \texttt{sleep} depends on some variables which are influenced by the data in the test case (and not just a constant in the SUT).
Having a fuzzing process in which each single test case timeouts would drastically reduce the number of fitness evaluations during the search, reducing the amount of the search landscape that can be explored.

Furthermore, even if a \texttt{sleep} is executed on a background thread, when the evaluation of a test case is completed, we would want to avoid having dangling sleeping threads that might suddenly wake up when evaluating a new test case later on.
Test case executions should be independent from each other.
As such, when the execution of a test case is completed, before evaluating a new test, the \emph{state} of the SUT should be reset.
A typical example is to clean the modifications applied on the connected databases (if any).
Dangling threads is yet another case of SUT state that needs to be handled.

We provide a method replacement to handle all the cases of \texttt{sleep}, with two main objectives:
(1) collect information on all threads that sleep, so they can be ``interrupted'' automatically once a test case execution is completed (also note that \evo is equipped with a ``kill-switch''~\cite{arcuri2023building} to block the thread execution as soon as the awoken thread executes any instrumented code);
(2) put a limit to the amount of time the thread is allowed to sleep (e.g., 1 second).

To avoid possibly messing up with the threads of the HTTP server of the SUT (e.g., Tomcat), sleeps in those threads are not modified.
Also, technically speaking, point (2) could change the behavior of the SUT.
Doing this is arguably controversial, but the benefits strongly outweigh the negative sides.
Leaving the sleeps as they are could just make the fuzzing unfeasible in some cases (due to drastic performance drops), and TCP timeouts together with ``dirty'' state between different test executions would be worse than possibly breaking some ``soft-constraint'' time-related behavior.
In other words, in the large majority of the cases when fuzzing REST APIs, we are not dealing with test executions that should last for minutes, where the SUT is still executing business code after it has responded to an incoming HTTP request.
In those special, rare cases, more sophisticated techniques would need to be designed.

\subsection{String Specializations}
\label{sub:uri}

Strings are widely used as input in web services.
It is one of the most common types, if not the most common one (although being able to claim that would require an analysis of existing APIs on internet).
However, strings are problematic for testing purposes.
Each string defines a hugely massive search space of possible combinations of characters.
Given $k$ possible characters, and length up to $n$ characters, there are $\sum_{0 to n}k^n$ possible strings.
During the search, only an extremely tiny subsets of all those possible strings can be evaluated.
Therefore, there is the need of smart strategies when dealing with string inputs.

Often, strings should match some specific constraints, like representing a valid email or a valid IP address.
Those constraints could be for example represented with regular expressions, which define a subset of valid strings for the given specialized string type.
When generating test inputs, it would make sense to generate valid strings based on those constraints.
Although, of course, for robustness testing it still makes sense to send some invalid strings any now and then.
Regular expressions are widely used, and so this common case was already handled in~\cite{arcuri2021enhancing} (e.g., for classes such as \texttt{java.util.regex.Pattern}).
However, there are several JDK classes that represent specialized strings, and that can take strings as input when initializing new instances of such objects.
Common examples are URIs (constructor for \texttt{java.net.URI}, and methods such as \texttt{URI.create()} and \texttt{URI.resolve()})
and URLs (e.g., constructor for \texttt{java.net.URL}).
When dealing with databases, Universal Unique Identifiers (UUIDs) are also common (e.g., the method \texttt{fromString()} in the class \texttt{java.util.UUID}).

We provide method replacements for all these methods, to enable taint analysis in them.
When in our instrumentation we detect that a tainted input is given as input to any of these replaced methods, the search is informed about it.
To enable generating valid strings that would not crash those methods (i.e., throw an exception due to invalid inputs), we extended the search engine in \evo to provide new specialized \emph{genes} for the evolving test cases.
In \evo, each gene not only defines the structure of the data (i.e., the \emph{genotype}) and how it will be represented when evaluating the fitness function (i.e., the \emph{phenotype}), but also it defines the mutation operator on such type (e.g., mutating an integer gene is different from mutating a string gene).

The simplest case is \texttt{UUIDGene}, which is a 128-bit label.
Internally, in its genotype it includes two \texttt{LongGene}s, which are then mutated like any \texttt{LongGene} in \evo.
The right phenotype is then reconstructed from using the constructor of \texttt{UUID} that takes as input two \texttt{long} inputs.

\begin{figure*}[t!]
\centering
\includegraphics[width=0.95\linewidth]{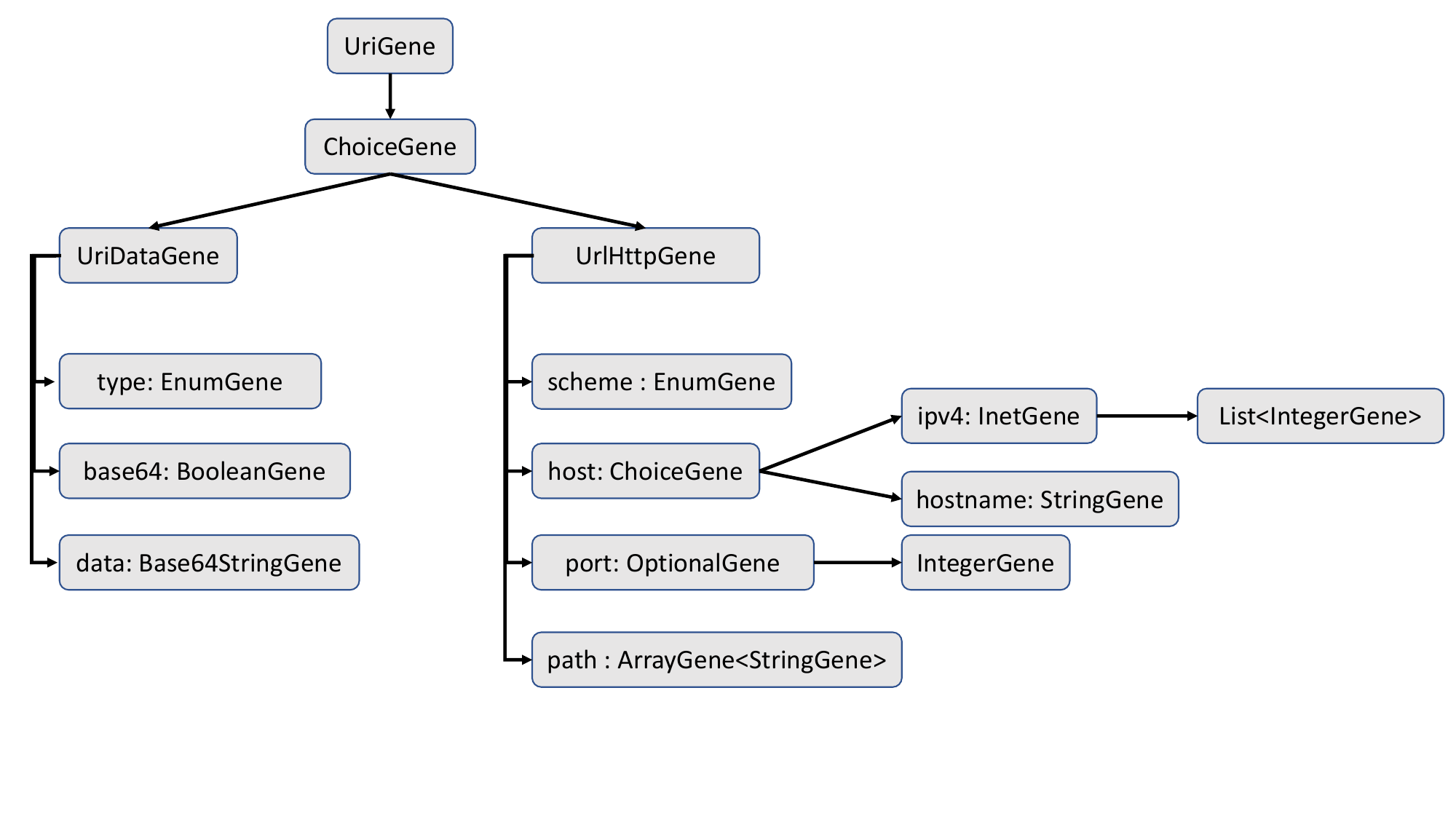}
\caption{
\label{fig:urigene}
Simplified tree-representation for new \texttt{UriGene} introduced to handle URI strings.}
\end{figure*}

The cases of URI and URL are similar, but unfortunately much more complex, as there are many rules on how to define valid URI/URL objects (e.g., see RFC 1738\footnote{https://www.rfc-editor.org/rfc/rfc1738}).
Figure~\ref{fig:urigene} shows a \emph{simplified} example of the tree-representation of the new \texttt{UriGene} we introduced for \evo.
Each valid URL is a valid URI, so a URI can be represented with a URL,  or a URN.
The syntax of a URI depends on its scheme, e.g.,
\texttt{data},
\texttt{http},
\texttt{https},
\texttt{file},
\texttt{ftp},
and
\texttt{gopher}.
So, first there is the need to make a ``choice'' of which scheme to use, which is done with a \texttt{ChoiceGene}.
In a \texttt{ChoiceGene}, only one of its children contributes to the phenotype of the individual, where the mutation operator selects which child to use.
For example, the \texttt{UriDataGene} used to represent the \texttt{data} protocol would have genes to define the type (e.g., an enumeration with entries like \texttt{text/plain}), whether it should be in base 64 format (using a \texttt{BooleanGene}), and the data itself.
The phenotype for such string specialization could create values such as \texttt{data:text/plain;base64,Zm9v}.
On the other hand, to represent an HTTP/S URL, we would need to define the scheme (e.g., an \texttt{EnumGene} with the values \texttt{http} and \texttt{https}), the host (which could either be string hostname or a numerical IP address), as well as an optional port, and path component.
Note: all these genes also have valid constraints (which are kept satisfied by the sampler and mutation operators), like for example the port gene is constrained in 0 and 65535, and each of the 4 integer genes inside the \texttt{InetGene} are constrained in 0 and 255.

\subsection{HTTP and JSON}
\label{sub:http}

In Section~\ref{sub:expansion} we have explained how we dealt with \emph{genotype expansion} in~\cite{arcuri2021enhancing} for handling underspecified schemas.
In particular, how to deal with missing query and header parameters, as well as non-specified types for body payloads.

What done here is a simple, straightforward extension of~\cite{arcuri2021enhancing}.
Where in~\cite{arcuri2021enhancing} we only handled the methods in the Spring interface \texttt{WebRequest}, here we also consider the JEE class \texttt{HttpServletRequest}, which has the same methods with same semantics, i.e., \texttt{getParameter()}, \texttt{getParameterValues()}, \texttt{getHeader()} and \texttt{getHeaders()}.
Furthermore, besides supporting GSON to analyze how strings are marshalled into JSON objects, now we do the same as well for Jackson (in particular, all the different variants of \texttt{readValue()} and \texttt{convertValue()})
With this, we now support all the major libraries for parsing JSON objects on the JVM.

\subsection{Javax/Jakarta Bean Validation}
\label{sub:validation}

Web APIs on the JVM are often implemented with enterprise frameworks, like for example Spring and JEE.
These frameworks make a large use of \emph{Bean} classes.
Those are classes that get \emph{enhanced} at runtime, based on annotations applied on them.
When the application starts, these frameworks instantiate \emph{proxy} classes, in which each method invocation is intercepted and possibly modified based on the semantics of the applied annotations (e.g., to automatically handle SQL transactions).

\begin{figure*}[t!]
	\centering
\begin{lstlisting}[language=java]
@Validated
@RestController
@RequestMapping(path = "/api/valid")
public class ValidRest {

    @RequestMapping(method = RequestMethod.POST)
    public String check(@RequestBody  @Valid ValidDto dto){
        if(dto == null){
            return "WRONG";
        }
        return "OK";
    }
}
\end{lstlisting}
\caption{
\label{fig:valid}
Example of bean definition for a Spring controller dealing with a REST endpoint.
}
\end{figure*}

Figure~\ref{fig:valid} shows such an example.
Here, the Spring framework treats the class \texttt{ValidRest} as a bean, as it is marked with the annotation \texttt{@RestController}.
The method \texttt{check()} will handle all \texttt{POST} requests for the endpoint \texttt{/api/valid} (this is specified with the annotation \texttt{@RequestMapping}).
The body payload of the incoming HTTP \texttt{POST} request is marshalled into an instance called \texttt{dto} of the class \texttt{ValidDto} (based on the annotation \texttt{@RequestBody}), which is then given as input to the method \texttt{check()}.
Because the class is marked with \texttt{@Validated} and then input \texttt{dto} is marked with \texttt{@Valid}, then all the constraints in such object are checked.
If any is violated, then the Spring framework will automatically return a failing HTTP response with status 400, i.e., user error, without executing \texttt{check()}.

To define constraints, JEE/Jakarta provides a rich system of annotations (used by Spring as well), such as
\texttt{@Min},
\texttt{@Max},
\texttt{@Positive},
\texttt{@PositiveOrZero},
\texttt{@Negative},
\texttt{@NegativeOrZero},
\texttt{@Size},
\texttt{@NotEmpty},
\texttt{@NotBlank},
\texttt{@Null},
\texttt{@NotNull},
\texttt{@AssertTrue},
\texttt{@AssertFalse},
\texttt{@Pattern},
and many more.
These annotations can then be applied on the fields of Java Beans, and validated each time the methods of the beans are called.
These annotations are not used only for REST APIs, but for all kinds of beans, including for example the beans used to represent data in databases (i.e., \texttt{@Entity} classes used in JPA, which we will discuss in more details in Section~\ref{sec:sql}).

Depending on the constraints in the annotated objects, this can be a major issue for automated testing purposes.
A HTTP call toward the endpoint would likely fail with a 400 status without any code of the business logic being executed.
If no code of the business logic is executed, then there would be no info on its code coverage and other heuristics such as the branch distance on its predicates (e.g., \texttt{if} statements).
The problem is that the check for constraints would be done in a call to \texttt{validate()} in the class \texttt{javax.validation.Validator}, deep inside the internals of the Spring framework.
Without any adhoc technique, even white-box fuzzers such as \evo would have no way to generate valid data for such endpoints (unless all the constraints are specified as well in the OpenAPI schema).
Note that the artificial example in Figure~\ref{fig:valid} is coming from one of our own end-to-end (i.e., system level) tests for \evo~\cite{arcuri2023building},
where the class \texttt{ValidDto} has 20 fields with constraint annotations.
Without any fitness gradient in the search, it would be very unlikely to sample a random \texttt{dto} instance that satisfies all these constraints.
However, with our novel technique presented in this paper, it becomes relatively simple, or at least simple enough that it can be solved consistently with a small search budget, and so it can be used as a end-to-end test for \evo itself~\cite{arcuri2023building}.

To handle these cases, we provide a method replacement for \texttt{Validator.validate()}.
Each time it is called, it computes two distances:
how far the object is from being evaluated as valid, and how far it is from being evaluated as invalid.
However, although there is only few places (typically just one) in which \texttt{Validator.validate()} is called (and where the transformation is applied), there could be hundreds of places in the business logic of the SUT that would trigger that validation check (e.g., all endpoint entry points, all intra-bean calls, and all write operations to the database when using JPA).
For the search, we need to be able to distinguish most, if not all, of these cases.
To achieve this, we create 2 testing targets (i.e., for the \texttt{true} and \texttt{false} outcome) for each
combination of endpoint name with HTTP verb and name of validated object.
In this example, those 2 new testing targets would be identified with something like
\texttt{VALIDATE\_POST:/api/valid\_ValidDto\_true} and \texttt{VALIDATE\_POST:/api/valid\_ValidDto\_false}.

Regarding the heuristic distance on the validated objects,
it can be considered as a conjunction of clauses, where each constrained field is a clause (e.g., $A \land B \land \dots$).
\emph{All} clauses must be satisfied for the object to be valid.
Note that fields could be objects as well, and so this needs to be applied recursively.
For the object to be considered invalid, we need only a single constraint to be evaluated as false, and so it can be represented with a disjunction of negated clauses (e.g., $\neg A \lor \neg B \lor \dots$).
Those cases can be handled with standard equations in the SBST literature~\cite{gallagher1997adtest}:
\begin{equation*}
d(A \land B) = d(A) + d(B)
\end{equation*}
\begin{equation*}
d(A \lor B) = min(d(A), d(B))
\end{equation*}
As the constraints are based on annotations, and not on executed methods that could have side-effects, here there is no issue when dealing with short-circuit evaluations of boolean predicates (which would require more advanced equations, as done for example in~\cite{zhang2023javascript}).

Most of these constraints in JEE/Jakarta deal with numbers and strings, which just require a rather straightforward mapping of existing branch distance calculations~\cite{Kor90,Alshraideh06} and use of taint-analysis (especially important for \texttt{@Pattern}).
We have handled all of those cases.
However, there are two groups of constraints that we do not handle yet.
First, we do not deal with any time-related constraints, like for example
\texttt{@Future},
\texttt{@FutureOrPresent},
\texttt{@Past},
and
\texttt{@PastOrPresent}.
Time related properties are hard to handle in automated testing, and can be a major source of \emph{flakiness}.
For example, a time constraint valid during the search might not be longer valid when the generated test cases are executed later on.
Without a proper, deterministic handling of time behaviours, trying to handle those constraints would not be particularly useful.
Second, we do not handle \emph{custom} constraints.
JEE/Jakarta enables users to write their own annotations with customized code to evaluate the validity of the constraints.
As such code could do anything, we cannot prepare pre-defined heuristic distances for those cases.
Future work would be needed to define on-the-fly heuristics based on the analysis of the source code of these custom methods.

\section{Underspecified REST API Schemas}
\label{sec:schemas}

\begin{figure}[!t]
\begin{lstlisting}[language=java,basicstyle=\footnotesize]
@RequestMapping(value="paypal/ipn/consumer/{consumerID}",
                method=RequestMethod.POST)
protected void consumerLoadUpConfirmation(
      @PathVariable(value = "consumerID") long cid,
      HttpServletRequest request,
      HttpServletResponse response
      ) throws ServletException, IOException {
  Map<String,String> configurationMap = Configuration.getConfig();
  IPNMessage ipnlistener = new IPNMessage(request, configurationMap);
  boolean isIpnVerified = ipnlistener.validate();
  String transactionType = ipnlistener.getTransactionType();
  Map<String,String> map = ipnlistener.getIpnMap();

  String payerEmail = map.get("payer_email");
  Double quantity = Double.valueOf(map.get("mc_gross"));
\end{lstlisting}
\caption{\label{fig:proxyprint:params}
Snippet of function handler for the endpoint \texttt{/paypal/ipn/consumer/\{consumerID\}} in \emph{proxyprint}.
}
\end{figure}

In~\cite{arcuri2021enhancing} we did present some initial work to handle under-specified OpenAPI schemas (recall Section~\ref{sub:expansion}), which has been now extended here by dealing with more classes and libraries (Section~\ref{sub:http}).
However, still such work would only be able to deal with Spring and JEE/Jakarta, and, even for those, not all cases could be handled.
Let us consider the snippet in Figure~\ref{fig:proxyprint:params}, coming from the SUT \emph{proxyprint} used in our empirical analysis (Section~\ref{sec:study}).
This is one of cases of \emph{open problems} that was discussed in details in~\cite{zhang2023open}:

\begin{quote}
\emph{
``...  shows an example in which the line defining the variable \texttt{quantity} throws an exception, due to \texttt{Double.valueOf} being called on a null input.
      Here, an HTTP object \texttt{request} is passed as input to the constructor of \texttt{IPNMessage}, which is part of PayPal SDK library.
      Inside such library, \texttt{request.getParameterMap()} is called to extract all the parameters of the HTTP request, which are used to populate the map object returned by \texttt{ipnlistener.getIpnMap()}.
      However, as such parameters are read dynamically at runtime, the OpenAPI/Swagger schema has no knowledge of them (as for this SUT the schema is created automatically with a library when the API starts).
      Therefore, there is no info to use an HTTP parameter called \texttt{mc\_gross} of type double''~\cite{zhang2023open}.
}
\end{quote}

Creating a method replacement for \texttt{request.getParameterMap()} would not help much here, as, at that point in time when it is called, there is no information yet on the parameter name \texttt{mc\_gross} that is going to be read later on in the SUT's business logic.
There is the need for a more general solution that is able to handle also these cases.

Our novel solution works as follows.
First, when we make HTTP calls, we add a ``fake'' HTTP header (e.g., \texttt{x-EMextraHeader123}) and a ``fake'' query parameter (e.g., \texttt{EMextraParam123}).
Then, each time any collection (e.g., lists, sets and maps) is queried with a key/value $X$, we check if such collection contains our extra header or parameter names.
All data structures are already instrumented (e.g., recall Section~\ref{sub:data-structures}) to enable heuristic computations and taint analysis, so this is just an extra check done inside those method replacements.
Then, if there is a match, we check if $X$ is an already known header/parameter name (e.g., from the OpenAPI schema definition).
If not, then there might be a good chance that we are in a case like in Figure~\ref{fig:proxyprint:params}.
The intuition here is that an HTTP server/framework would read all incoming headers/params and put them in a data structure, and then check for specific names in such data structure based on what the business logic of the SUT needs (e.g., \texttt{mc\_gross} in this case).
If that happens, then at the next fitness evaluation, we replace the name of the fake header/param with $X$, with a randomly initialized or tainted value.
Then, with taint-analysis on the input of \texttt{Double.valueOf} we can further infer in the following fitness evaluation that the value of $X$ should be turned into a double.
Besides dealing with method calls on collections (e.g., search for a key in a map), we also check for these extra params/headers in every string comparisons (e.g., in \texttt{String.equals()}).

This approach is fully automated, and it does solve the problem like in the example in Figure~\ref{fig:proxyprint:params}.
However, there are two important aspects to consider here.
First, it is performance.
Sending extra query parameters and headers has a negligible cost, but checking for string matching might not, especially not on large collections.
This cost can be worthy if the OpenAPI schema is underspecified.
However, it would be a clear performance loss if it is not.
But we cannot really know if there is any missing header or parameter in the schema before applying our technique.
What can be done is to try to minimize its computational cost.
We apply two simple heuristics.
First, the sending of extra parameters and headers is done only for a short period of time (e.g., currently 10\% of the search budget, like for the first 6 minutes if the search budget is 1 hour).
Second, if the analyzed collection is too large (e.g., more than 16 elements), then it would be unlikely that it represents the storing of HTTP headers or query parameters.
If so, in these cases we simply skip any string matching checks in the instrumented method replacements.

Another rather peculiar issue is that, as we found out during some preliminary experiments, there might be some special headers automatically handled by the enterprise framework used by the API.
For example, in older versions of Spring the HTTP filter \texttt{HiddenHttpMethodFilter} was active by default.
This would handle a special query parameter named \texttt{\_method}, to enable to change the HTTP verb of incoming requests.
For example, a user could make a \texttt{POST} HTTP request with a query part \texttt{?\_method=PUT} to tell Spring to consider the request as a \texttt{PUT}.
This feature comes from old time requirements, before the widespread use of JavaScript and AJAX, when it was possible to do only \texttt{GET} and \texttt{POST} requests from a HTML page (e.g., by clicking on a \texttt{<a>} link or submitting a \texttt{<form>}).
What happened here is that our technique would automatically discover such hidden parameter \texttt{\_method}, and use it in the evolving test cases.
This provides no benefit to the search, and actually just lead to plenty of useless tests that return a 405 HTTP status code (``Method Not Allowed''), e.g., when sending random strings such as \texttt{?\_method=Pfre5}.
The solution here is to simply ignore any newly discovered parameter that is named  \texttt{\_method}.

\section{Underspecified SQL Constraint Schemas}
\label{sec:sql}

Data in SQL database can have constraints, which are checked each time a new entry is added.
Constraints could be as simple as stating that  the entries of a column are unique, or that they might or might not be nullable.
However, there might more sophisticated cases, in which custom constraints can be written with \texttt{CHECK} directives in the SQL table schema.

For a white-box fuzzer, it is important to be able to create data directly into the SQL databases as part of the fuzzing process.
This is needed to be able to test scenarios when the same SQL database is used by more than one application (e.g., in producer/consumer scenarios, or when data is populated by external batch jobs and the tested API is read-only).
Also, it might improve the readability of the generated tests if some specific needed data in the SQL database could only be obtained with a long sequence of HTTP calls to the API.
For this reason, \evo is able to automatically generate data into SQL databases as part of the search~\cite{arcuri2020sql}.
During the search, it analyzes all executed \texttt{SELECT} commands, to see which ones return no data.
In those cases, it will automatically add data in the queried tables, with the fitness function aiming at solving the constraints in the \texttt{WHERE} clauses of those empty \texttt{SELECT}s~\cite{arcuri2020sql}.
Each test case will then be enhanced with extra initializing actions for the database, whose inputs (i.e., the data to insert) will be evolved during the search in the same way as any other element of the HTTP requests~\cite{arcuri2020sql}.

\begin{figure*}[!t]
\begin{lstlisting}[language=java,basicstyle=\footnotesize]
@Entity
@Table(name = "app_session")
public class VerificationAppSession implements Serializable {

  private static final long serialVersionUID = 1L;

  @Id
  @GeneratedValue(strategy = GenerationType.IDENTITY)
  @Column(name = "id")
  private Long id;

  @Version
  @Column(name = "version")
  private long version;

  @Column(name = "created_at")
  private LocalDateTime createdAt;

  @Column(name = "updated_at")
  private LocalDateTime updatedAt;

  @Column(name = "hashed_guid")
  private String hashedGuid;

  @Column(name = "hashed_guid_dob")
  private String hashedGuidDob;

  @Column(name = "registration_token_hash")
  private String registrationTokenHash;

  @Column(name = "tele_tan_hash")
  private String teleTanHash;

  @Column(name = "tan_counter")
  private int tanCounter;

  @Column(name = "sot")
  @Enumerated(EnumType.STRING)
  private AppSessionSourceOfTrust sourceOfTrust;

  @Column(name = "teletan_type")
  @Enumerated(EnumType.STRING)
  private TeleTanType teleTanType;
\end{lstlisting}
\caption{
\label{fig:appsession}
Snippet of the JPA entity class \texttt{VerificationAppSession} in the SUT \emph{cwa-verification-server}.
}
\end{figure*}

As adding invalid data would be pointless, as the SQL database would just straight-up reject them, when we insert SQL data in \evo we make sure that all (linear) constraints are satisfied.
However, there might be cases in which this is not enough.
Figure~\ref{fig:appsession} shows a snippet from the class \texttt{VerificationAppSession} in the SUT \emph{cwa-verification-server} in our case study (Section~\ref{sec:study}).
This class is marked with the JPA \texttt{@Entity} notation, and it is used to map data from the SQL table \texttt{app\_session}.
One problem here is that the numeric variable \texttt{tan\_counter} in that table is nullable, i.e., \texttt{NULL} is a valid value for it.
However, that entity class declares it with the primitive Java type \texttt{int}, instead of the nullable Java type \texttt{Integer}.
If there is a null value for that variable, the parsing of that entity would crash, throwing an exception.
Regardless of any JEE/Jakarta constrain annotation (e.g., \texttt{@Min} and \texttt{@Max}), primitive vs.~nullable types do implicitly define further constraints.
In this particular case, there is a mismatch of constraints between the SQL database and its ORM representation in the API.
This is a fault, but whether the fault is in the API (i.e., wrong constrain mapping) or in the SQL database definition (i.e., underspecified constraints) is something that only the authors of that API can really comment on.

Another issue can be seen for columns such as \texttt{teletan\_type}.
In the database, it is declared as a \texttt{varchar(10)}, i.e., a string of at most 10 characters.
However, the JPA entity defines a further constraint that such string is actually an enumeration, where only a limited set of values is acceptable.
In particular, the enumeration \texttt{TeleTanType} contains only the values \texttt{TEST} and \texttt{EVENT}.
A random string in that column would crash the JPA parsing of entries from such table.

Our solution for this problem is as follows.
First, every single time a class is loaded into the JVM, we check if it is a JPA entity (e.g., by checking for the annotation \texttt{@Entity}).
If so, we analyze all JEE/Jakarta constraints defined on it (recall Section~\ref{sub:validation}), including any implicit nullability check based on primitive types, and any enum declarations.
Then we compare these sets of constraints with the constraints derived directly from querying the SQL database~\cite{arcuri2020sql}.
For this, we need to identify the corresponding table in the SQL database.
We follow the same algorithms as done in JPA implementations such as Hibernate to resolve table and column names, which might be the same as the Java class/field names, or overridden with annotations, like \texttt{VerificationAppSession} vs.~\texttt{app\_session} (see Figure~\ref{fig:appsession}).
If the JPA entity provides more constraints, then we use those constraints when adding data into the database.
For example, we would not add a \texttt{NULL} into the column \texttt{tan\_counter}.
However, with a small probability, for robustness testing any now and then we also add data that does not satisfy these extra constraints.

\section{Timed Events}
\label{sec:timed_events}

In a backend system, it is possible to schedule tasks at precise time intervals.
For example, in the API \catwatch used in our case study (Section~\ref{sec:study}) a task is executed at 8:01am each day,
to fetch project data from GitHub.
In the API \ocvnrest data is imported each day at 3am and 9pm.
These background tasks are set with Spring annotations such as \texttt{@Scheduled}, and are executed independently from the HTTP requests to the API.

These background tasks introduce few complications for testing.
First, generated tests could become \emph{flaky} if the background tasks have side-effects (e.g., adding or deleting records from a database).
Second, it might invalidate empirical comparisons of fuzzers (and their different settings) when measuring code coverage metrics.
It is not uncommon that experiments for this kind of systems take days to run.
For example, when running experiments with a fuzzer $X$ at 2:59am on \ocvnrest, we might get better results than a fuzzer $Y$ at 4am simply because the code executed by the background task would contribute to the measured code coverage.

This is a particular problem for black-box fuzzers, as there is no programmatic way (at least in Spring) to disable all the scheduled tasks when starting the API for testing.
The API would need to be \emph{manually modified} to enable disabling background tasks during fuzzing experiments.
To handle this threat to the validity of the experiments, with a white-box approach using bytecode instrumentation we can simply remove any \texttt{@Scheduled} annotations when classes are loaded into the JVM (and so the tasks are not executed).

\section{Empirical Study}
\label{sec:study}

In this paper, we have carried out an empirical study to answer the following research questions:

\begin{enumerate}

\item[{\bf RQ1}:] What is the impact on line coverage and fault detection of our novel white-box techniques?

\item[{\bf RQ2}:] How do the heuristic computation overhead and search budget correlate?

\end{enumerate}

\subsection{Case Study}
\label{sub:case-study}

\begin{table*}
	\centering
	\caption{Statistics of the used SUTs in the empirical study.}
	\label{tab:suts}
		\begin{tabular}{l rrr}\\ 
\toprule 
SUT & \#SourceFiles & \#LOCs & \#Enbdpoints \\
\midrule
\emph{catwatch} &  106 &  9636 &  14 \\
\emph{cwa-verification} &  47 &  3955 &  5 \\
\emph{features-service} &  39 &  2275 &  18 \\
\emph{genome-nexus} &  405 &  30004 &  23 \\
\emph{gestaohospital} &  33 &  3506 &  20 \\
\emph{ind0} &  103 &  17039 &  20 \\
\emph{languagetool} &  1385 &  174781 &  2 \\
\emph{market} &  124 &  9861 &  13 \\
\emph{ocvn} &  526 &  45521 &  258 \\
\emph{proxyprint} &  73 &  8338 &  74 \\
\emph{rest-ncs} &  9 &  605 &  6 \\
\emph{rest-news} &  11 &  857 &  7 \\
\emph{rest-scs} &  13 &  862 &  11 \\
\emph{restcountries} &  24 &  1977 &  22 \\
\emph{scout-api} &  93 &  9736 &  49 \\
\midrule
Total 15 & 2991 & 318953 & 542 \\
\bottomrule 
\end{tabular} 

\end{table*}

To evaluate our novel techniques, we used all the 14 RESTful APIs present in the EMB corpus~\cite{icst2023emb} at the time of these experiments, in particular version 1.6.1~\cite{zenodo161emb}.
EMB is a corpus of Web APIs (including GraphQL and RPC APIs), which we have collected and extended each year with new APIs since 2017.
EMB includes the \evo drivers for enabling white-box fuzzing on all of these APIs.
To obtain more generable results, besides using open-source APIs, we also included in our experiments one API from one of our industrial partners.
For this paper, to refer to this API we use the fictional name \emph{ind0}.
Table~\ref{tab:suts} shows some statistics on these 15 RESTful APIs,
including number of source files, lines of code and number of REST endpoints in each of these APIs.
Note these code statistics count only what present in business logic of those APIs.
Statistics on code in third-party libraries (e.g., HTTP servers) are not included here.

EMB provides APIs of different size and complexity, coming from different domains, covering a variegated set of APIs needed for scientific experimentation~\cite{icst2023emb}.
Some are artificial APIs aimed at studying how to deal with numeric (\emph{rest-ncs}) and string (\emph{rest-scs}) constraints.
Others come from public administrations (e.g., \emph{ocvn}),
or are widely popular tools that provide a REST interface (e.g., \emph{languagetool}).
A full description of these APIs can be found at~\cite{icst2023emb,EMB}.

\subsection{Experiment Settings}
\label{sub:settings}

In this paper, we have carried out two different sets of experiments.
In the first set, we considered and compared 6 different configurations of \evo,
namely:

\begin{description}
\item[Base]: default version of \evo, without any of our novel techniques, besides what presented in Section~\ref{sec:timed_events}.
            This latter is on in the \emph{Base} version as well because it can impact the fairness and soundness of the comparisons.
\item[TAOS]: short for ``Taint Analysis on Sampling'', in which we use tainted values (with a certain probability, e.g., 90\%) when test cases are sampled, and not just when test cases are mutated~\cite{arcuri2021enhancing}.
This is a rather minor modification to the MIO algorithm~\cite{arcuri2018test}, but turned out that can have quite an impact on results.
\item[TT]: all the testability transformations presented in Section~\ref{sec:new_replacements} are activated.
\item[TT+OpenAPI]: configuration TT plus the handling of underspecified OpenAPI schemas presented in Section~\ref{sec:schemas}.
    Note that the techniques presented in Section~\ref{sec:schemas} relying on instrumentation for JVM collections, and so that is why it requires TT on.
    Disabling all the other TT heuristics just for the sake of these experiments would had required significant engineering effort, which we did not considered worthy to invest.
\item[JPA]: handling of underspecified SQL schemas presented in Section~\ref{sec:sql}.
\item[All]: all new techniques presented in this paper are activated at the same time.
\end{description}

Fuzzing sessions were run for 1 hour each.
To take into account the randomness of search-based fuzzing,
each experiment was repeated 10 times.
In total, considering 6 configurations and 15 SUTs, this required
$6 \times 15 \times 10 = 900$ hours, i.e., $37.5$ days of computation.
To be able to run all these experiments in reasonable time, they were run in parallel (15 at a time) on a
HP Z6 G4 Workstation with Intel(R) Xeon(R) Gold 6240R CPU @2.40GHz 2.39GHz, 192 GB RAM, with 64-bit Windows 10 OS.

Ideally, the impact of each single new technique (e.g., each single testability transformation presented in Section~\ref{sec:new_replacements}) should be studied in isolation.
However, that would have a non-trivial engineering cost (e.g., to 
support enabling only specific subsets of testability transformations in \evo's code instrumentator), as well as making the running of all the experiments unfeasible in reasonable time.
Using 6 configurations for the experiments was a viable compromise.

Still, it is important to check that each single technique is useful in its own right.
\evo is an industry-ready tool (e.g., used daily in large companies such as Meituan on hundreds of microservices~\cite{meituanArxiv2022,zhang2023rpc}), and particular care is taken to verify the correctness of its components.
For this goal, \evo has a sophisticated system of end-to-end tests, where it is run on a set of artificial API examples, carefully crafted to study and verify each single of its features~\cite{arcuri2023building}.
These tests are automatically run in CI (e.g., GitHub Actions).
To verify the correctness of our novel techniques, we created artificial APIs and new end-to-end tests for all of them.
An example is \texttt{ValidEMTest},\footnote{\url{https://github.com/EMResearch/EvoMaster/blob/master/e2e-tests/spring-rest-openapi-v2/src/test/java/org/evomaster/e2etests/spring/examples/valid/ValidEMTest.java}}
used to verify the handling of bean validation presented in Section~\ref{sub:validation}.
All these new system level tests pass, and are currently part of the daily CI testing of \evo.
As discussing all these end-to-end tests would take considerable amount of space, we refer the interested reader to the code repository of \evo (e.g., the module \texttt{e2e-tests}).

Based on the results of these experiments, after analyzing its results, to get more insight on a potential issue, we carried out a second set of experiments using a single SUT, namely \emph{ocvn}.
In this case, fuzzing sessions were run for 10 hours instead of just 1 hour.
But only two settings were considered, \emph{Base} and \emph{All}, with 10 repetitions.
This took
$10 * 10 * 2 = 200$ hours, i.e., $8.3$ days.
Experiments were run on the same hardware and configurations of the first set of experiments.

In total, the computation cost of our experiments is $37.5 + 8.3 = 45.8$ days.
Note that, although experiments can be run in parallel, there is a limit on how many can be parallelized, considering the specification of the employed hardware.
For example, each single experiment requires to run few processes, like \evo itself, the tested API and potentially its databases (e.g., Postgres through Docker) if any is in use.
This can take significant amount of OS resources, such as RAM and OS-level threads.
Overloading the target machine with too many experiments in parallel could have negative impact on time-based comparisons.

\subsection{Results for RQ1}
\label{sub:rq1}

\begin{table*}
	\centering
	\caption{
	Performance comparisons between the \emph{Base} and \emph{All} configurations, in terms of average (i.e., arithmetic mean) line coverage and average number of detected faults.
	Results of statistical tests are reported, including p-values and $\hat{A}_{12}$ effect sizes.
	For p-values lower than the threshold $\alpha=0.05$, the effect sizes $\hat{A}_{12}$  are shown in bold.
	We also report the average number of HTTP calls done during the search, and their scaled difference compared to the \emph{Base} configuration, i.e., $\frac{avg(Base)-avg(All)}{avg(Base)}$.
	}
	\label{tab:all}
	\resizebox{1.\linewidth}{!}{
		\begin{tabular}{ l rrrr rrrr rrr}\\ 
\toprule 
SUT & \multicolumn{4}{c}{Line Coverage \%} & \multicolumn{4}{c}{\# Detected Faults} & \multicolumn{3}{c}{\# HTTP Calls} \\ 
    & Base & All  & $\hat{A}_{12}$ & p-value  & Base & All & $\hat{A}_{12}$ & p-value & Base & All & Difference \\ 
\midrule 
\emph{catwatch} & 42.5 & 47.1 & {\bf 0.97} & $< 0.001$ & 19.4 & 25.2 & {\bf 1.00} & $< 0.001$ & 3120 & 14352 & -360.04\\
\emph{cwa-verification} & 47.4 & 57.6 & {\bf 1.00} & 0.010 & 7.4 & 12.0 & {\bf 1.00} & 0.010 & 151478 & 122480 & +19.14\\
\emph{features-service} & 81.5 & 81.5 & 0.39 & 0.396 & 33.2 & 34.1 & 0.61 & 0.424 & 182534 & 184688 & -1.18\\
\emph{genome-nexus} & 36.7 & 36.5 & 0.44 & 0.705 & 20.0 & 21.0 & 0.66 & 0.216 & 59304 & 41742 & +29.61\\
\emph{gestaohospital-rest} & 39.6 & 39.6 & 0.50 & 1.000 & 22.0 & 22.0 & 0.50 & 1.000 & 240280 & 229379 & +4.54\\
\emph{ind0} & 12.4 & 23.1 & {\bf 1.00} & $< 0.001$ & 44.1 & 59.1 & {\bf 1.00} & $< 0.001$ & 232335 & 177159 & +23.75\\
\emph{languagetool} & 41.7 & 40.0 & 0.29 & 0.146 & 8.3 & 11.2 & 0.75 & 0.080 & 13713 & 21742 & -58.55\\
\emph{market} & 48.6 & 47.5 & 0.28 & 0.117 & 20.0 & 20.2 & 0.54 & 0.781 & 27731 & 27270 & +1.66\\
\emph{ocvn-rest} & 37.1 & 37.1 & {\bf 0.84} & 0.008 & 553.8 & 500.0 & {\bf 0.00} & $< 0.001$ & 136751 & 103057 & +24.64\\
\emph{proxyprint} & 53.2 & 53.7 & 0.57 & 0.633 & 83.6 & 84.7 & 0.56 & 0.688 & 40095 & 38105 & +4.96\\
\emph{rest-ncs} & 93.0 & 93.0 & 0.50 & 1.000 & 6.0 & 6.0 & 0.50 & 1.000 & 275741 & 262458 & +4.82\\
\emph{rest-news} & 66.9 & 67.7 & {\bf 0.82} & 0.008 & 8.0 & 7.8 & 0.40 & 0.167 & 300261 & 288004 & +4.08\\
\emph{rest-scs} & 85.7 & 86.0 & 0.61 & 0.437 & 12.0 & 11.9 & 0.45 & 0.368 & 283604 & 261644 & +7.74\\
\emph{restcountries} & 77.0 & 77.1 & {\bf 0.75} & 0.014 & 2.0 & 2.0 & 0.50 & 1.000 & 233799 & 224224 & +4.10\\
\emph{scout-api} & 52.9 & 53.4 & 0.58 & 0.567 & 89.5 & 88.1 & 0.39 & 0.437 & 157931 & 135622 & +14.13\\
\midrule 
Average  & 54.4 & 56.1 & 0.64 &  & 62.0 & 60.4 & 0.59 &  & 155912 & 142128 & -18.44\\
Median  & 48.6 & 53.4 & 0.58 &  & 20.0 & 21.0 & 0.54 &  & 157931 & 135622 & +4.82\\
\bottomrule 
\end{tabular} 

	}
\end{table*}

\begin{table*}
	\centering
	\caption{For each of the six analyzed configuration, we report average line coverage and average number of detected faults.
	Results that are statistically different from \emph{Base} (at $\alpha=0.05$ level) are reported in bold.}
	\label{tab:six}
	\resizebox{1.\linewidth}{!}{
		\begin{tabular}{ l rl rl rl rl rl rl}\\ 
\toprule 
SUT & \multicolumn{2}{c}{ Base }&\multicolumn{2}{c}{ TAOS }&\multicolumn{2}{c}{ TT }&\multicolumn{2}{c}{ TT+OpenAPI }&\multicolumn{2}{c}{ JPA }&\multicolumn{2}{c}{ All }   \\  
\midrule 
\emph{catwatch} & 42.5&19.4 & 41.9&18.7 & {\bf 45.7}&{\bf 25.3} & {\bf 47.9}&{\bf 26.2} & {\bf 45.1}&19.5 & {\bf 47.1}&{\bf 25.2} \\ 
\midrule 
\emph{cwa-verification} & 47.4&7.4 & 47.5&7.4 & 47.6&8.2 & {\bf 47.6}&8.0 & {\bf 57.0}&{\bf 10.2} & {\bf 57.6}&{\bf 12.0} \\ 
\midrule 
\emph{features-service} & 81.5&33.2 & 81.5&33.9 & 81.5&34.8 & 81.4&33.2 & 81.8&{\bf 35.5} & 81.5&34.1 \\ 
\midrule 
\emph{genome-nexus} & 36.7&20.0 & 37.3&20.6 & 36.9&20.9 & 36.4&20.5 & 36.3&20.1 & 36.5&21.0 \\ 
\midrule 
\emph{gestaohospital-rest} & 39.6&22.0 & 39.5&22.0 & 39.6&22.0 & 39.6&22.0 & {\bf 39.4}&22.0 & 39.6&22.0 \\ 
\midrule 
\emph{ind0} & 12.4&44.1 & {\bf 24.1}&{\bf 54.7} & 15.6&{\bf 50.6} & 13.5&{\bf 49.1} & 13.9&45.0 & {\bf 23.1}&{\bf 59.1} \\ 
\midrule 
\emph{languagetool} & 41.7&8.3 & 41.0&{\bf 5.4} & 41.1&7.4 & 41.9&11.2 & 38.3&7.9 & 40.0&11.2 \\ 
\midrule 
\emph{market} & 48.6&20.0 & 47.6&19.6 & 47.8&19.3 & 47.2&19.9 & 47.2&19.4 & 47.5&20.2 \\ 
\midrule 
\emph{ocvn-rest} & 37.1&553.8 & 37.1&{\bf 515.3} & 37.1&{\bf 539.9} & {\bf 37.1}&547.0 & 37.1&550.7 & {\bf 37.1}&{\bf 500.0} \\ 
\midrule 
\emph{proxyprint} & 53.2&83.6 & 52.9&86.9 & 51.5&82.6 & 54.1&86.2 & 54.0&82.1 & 53.7&84.7 \\ 
\midrule 
\emph{rest-ncs} & 93.0&6.0 & 93.0&6.0 & 93.0&6.0 & 93.0&6.0 & 93.0&6.0 & 93.0&6.0 \\ 
\midrule 
\emph{rest-news} & 66.9&8.0 & 67.4&8.0 & {\bf 67.6}&7.8 & 67.3&7.8 & 66.8&{\bf 7.6} & {\bf 67.7}&7.8 \\ 
\midrule 
\emph{rest-scs} & 85.7&12.0 & 85.7&11.9 & 86.0&12.0 & 86.2&12.0 & 86.3&12.0 & 86.0&11.9 \\ 
\midrule 
\emph{restcountries} & 77.0&2.0 & {\bf 77.1}&2.0 & 77.1&2.0 & 77.0&2.0 & 77.0&2.0 & {\bf 77.1}&2.0 \\ 
\midrule 
\emph{scout-api} & 52.9&89.5 & 52.7&89.3 & 53.0&90.6 & 53.2&89.4 & 54.4&90.1 & 53.4&88.1 \\ 
\bottomrule 
\end{tabular} 

	}
\end{table*}

Table~\ref{tab:all} shows the results in details for \emph{All} configuration compared to \emph{Base}.
We follow the statistical guidelines from~\cite{Hitchhiker14}, reporting $p$-values of Mann-Whitney-Wilcoxon U tests and Vargha-Delaney standarized $\hat{A}_{12}$ effect sizes.
Results are compared in term of line coverage and detected faults.
Detected faults are based on 500 HTTP status codes (server errors) and mismatches of the responses from the OpenAPI schemas for each distinct endpoint (more details on what kind of faults can be detected this way can be found in~\cite{marculescu2022faults}).
Note that there exist also other kinds of metrics that could be used for comparisons, such as for example branch and path coverage.
For sake of simplicity and space, we just report the two most common metrics used by practitioners in industry.

As our novel technique could introduce some non-trivial computation overhead in the fitness function, in Table~\ref{tab:all} we also report the number of HTTP calls done during the search.
This could give some insight on the computational cost of our techniques.
The less efficient (i.e., more computationally expensive) a technique is, the fewer number of HTTP calls can be done during the search budget (e.g., 1 hour in our case).
However, this needs to be analyzed with care, as the cost of each HTTP call is not the same.
For example, a call that returns immediately a 400 status code (user error) due to an invalid input would be faster than a successful 200 code call that executes large parts of the API's code, including interactions with external databases.
Better heuristics that lead to cover more code might result in executing HTTP calls that are more expensive to run, regardless of computation cost of the heuristics themselves.

To see the results of each of the 6 configurations in isolation, average results are summarized in Table~\ref{tab:six}.
Detailed statistical comparisons of each configuration with \emph{Base} are reported in Appendix.

From these results,
for branch coverage we can see an average improvement of $+1.7$\% (median $+4.8$\%) over the 15 APIs.
Results are statistical significant for only 6 APIs, with no statistically worse results.
On these APIs improvement are either
``small'' (e.g., less than 1\% for \emph{ocvn}, \emph{rest-news} and \emph{restcountries}),
``medium'' ($+4.6$\% for \emph{catwatch}),
or ``large'' (i.e., $+10.2$\% for \emph{cwa-verification} and $+10.7$\% for \emph{ind0}).
Note that the terms used here small, medium and large are subjective, and so technically arbitrary.

Regarding fault detection, statistically better results are found in 3 APIs.
However, there are statistically \emph{worse} results for 1 API, namely \emph{ocvn}.
There is large number of detected faults in \emph{ocvn} (e.g., more than 500), which is related to the large number of endpoints in this API (i.e., 258, recall Table~\ref{tab:suts}).
This outlier significantly impacts the statistics over the whole 15 APIs.
Although the average decreases from  $62.0$ to $60.4$, the median increases from $20.0$ to  $21.0$.
This is a special case in which the average value is roughly three times the median one.

We can make few hypotheses on why code coverage was significantly improved in only 6 out of 15 APIs.
\begin{itemize}
\item The case of \emph{rest-ncs} is easy to explain, as all maximum achievable coverage is already obtained~\cite{zhang2023open}, and so no further improvement is technically possible.
Running experiments on a ``solved'' SUT such as \emph{rest-ncs} is still valuable, for example to check if a novel technique does not make performance worse.
\item New code heuristics do have an impact only if the code in which they are applied is executed by the tests.
If that code is never reached, those new heuristics have no way to contribute to better performance.
A possible example of this phenomenon is  \emph{gestaohospital}, where achieved line coverage is $39.6$\%.
All of its missed branches seem related to interactions with a MongoDB database~\cite{zhang2023open}, for which currently \evo has no heuristics.
Once new techniques and heuristics to handle MongoDB databases are designed and implemented, more code of this API would be covered.
This new reached code might contain structures (e.g., branch statements) for which our novel heuristics could be helpful, and so improve performance even further.
\item   Even if a novel heuristic provides better gradient for the search, its computational cost might be non negligible, which might lead to fewer fitness evaluations.
        Those 2 contrasting phenomena might balance themselves out.
        A possible example here is \emph{genome-nexus}, where although code coverage stays very similar (with no statistically significant different), the number of HTTP calls is drastically reduced, from 59k to 41k.
\item Improvements could be applicable only on small part of the API's code.
      For example, an heuristic to better handle URL strings might have only a small impact on coverage if there is only a single call to \texttt{new URL(x)} in a code base of thousands of lines of code.
      Small improvements could be masked by the variance of the randomized process, and so be hard to detect.
      There are still many open problems in white-box fuzzing of Web APIs~\cite{zhang2023open}, and not all of them have same impact among all APIs.
      However, you could have a single branch that, once its constraints are solved, could lead to the execution of further thousands of lines of code (e.g., if related to input validation done at the beginning of HTTP call evaluation).
      This is hard to determine before running experiments.
\item  When presenting a set of new techniques, e.g., $X$ and $Y$, they might have conflicting side effects among them.
        For example $X$ could improve performance whereas $Y$ could reduce it, depending on the API.
        Studying both at same time might mask out the benefits provided by $X$.
        An example of this is \emph{catwatch} in Table~\ref{tab:six}, where each single technique improves performance but TAOS (although this latter is not statistically significant).
        When all are combined together, performance seems worse than just using TT+OpenAPI.
        The case of TAOS is quite peculiar.
        It gives drastic improvements to \emph{ind0}, nearly doubling the achieved code coverage, from $12.4$\% to $24.1$\%.
        However, it also significantly reduces the number of detected faults in \emph{ocvn}.
        Applying a new technique $Y$ only based on properties of a SUT could lead to address this issue, although how to determine which properties to use to estimate the impact of $Y$ does not seem trivial.
\end{itemize}

\begin{result}
{\bf RQ1}: Statistically better results were achieved on 6 out of 15 APIs, with average line coverage improvements up to 10.7\%.
With a search budget of 1 hour, statistically worse results were obtained for fault detection in 1 API, namely \emph{ocvn}.
\end{result}

\subsection{Results for RQ2}
\label{sub:rq2}

\begin{figure*}
\centering
\includegraphics[width=0.45\linewidth]{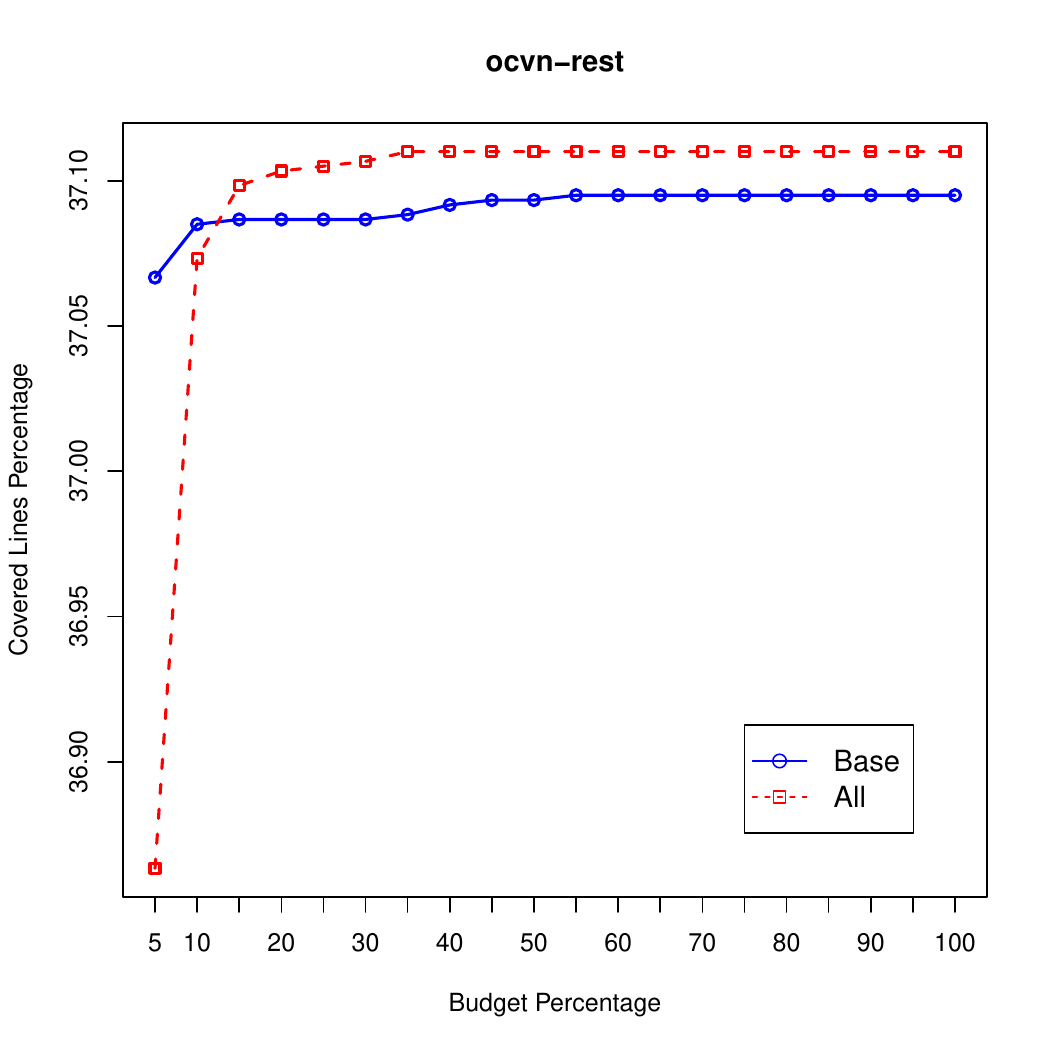}
\includegraphics[width=0.45\linewidth]{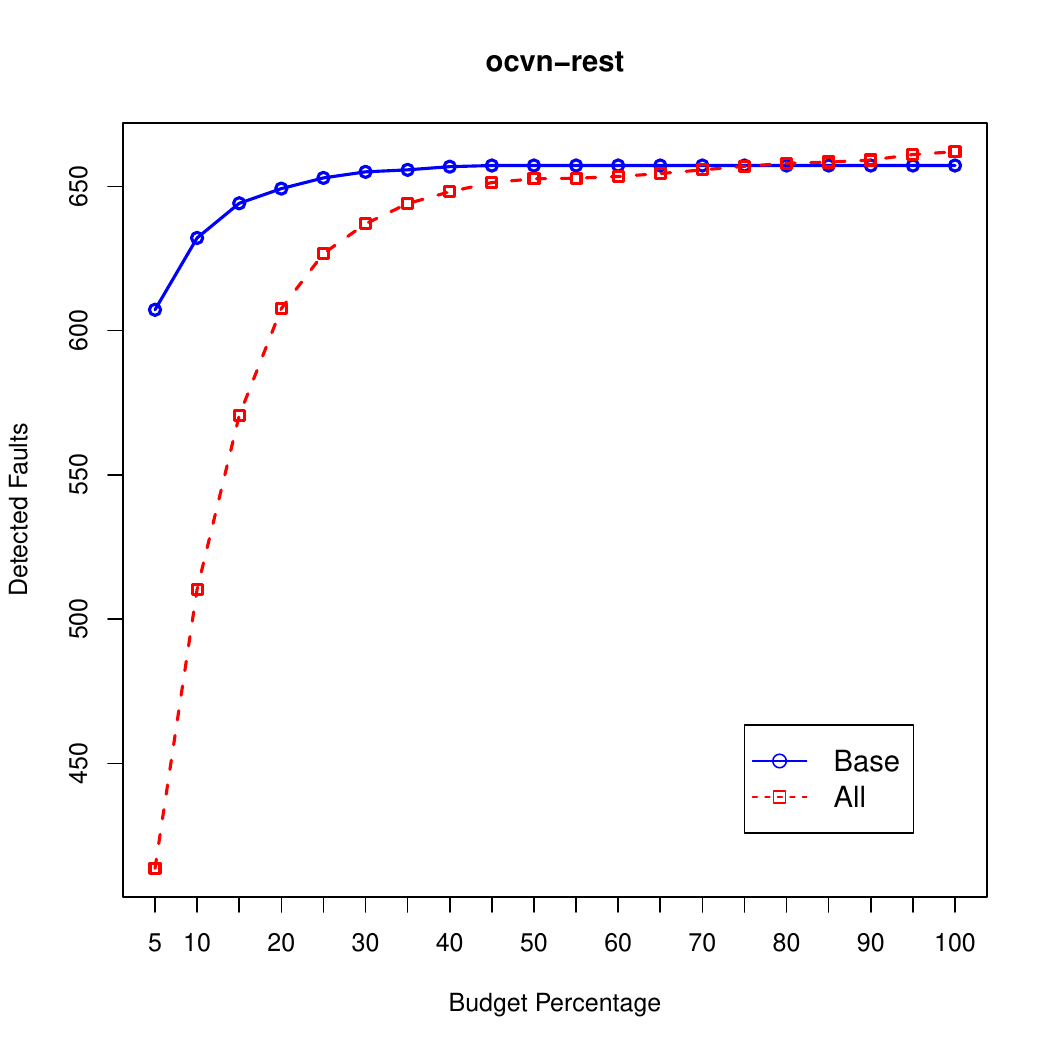}
\caption{
\label{fig:ocvn}
Performance of \emph{Base} and \emph{All} configurations on \emph{ocvn}, with search budget of 10 hours.
Results are displayed for each 5\% intervals of the search (i.e., every 30 minutes).
}
\end{figure*}

In the first set of experiments, our novel techniques achieved better results but in one case, namely \emph{ocvn}.
The computation cost of our novel techniques is not negligible on this API,
as the number of HTTP calls is reduced from 136k to 103k.
The question here is what would happen if the fuzzing would be run for longer.
The choice of 1 hour for the experiments is technically arbitrary, as arbitrary as 42 minutes or 24 hours.

After running for a certain amount of time, a search algorithm will be stuck on an \emph{optimum} (either \emph{global} or more likely a \emph{local} one).
Running the search for longer would not help much if the algorithm cannot escape from the local optima.
The search performance would reach a so called \emph{plateau}, and stagnates.
There is decades of research effort on addressing this problem,
e.g., for example with techniques such as \emph{fitness sharing}~\cite{sareni1998fitness} to increase population diversity in population-based evolutionary algorithms.
The tradeoffs between \emph{exploration} and \emph{exploitation} of the search landscape applied by a search algorithm do impact how, when and what type of local optimum would be reached.
In this regard, the MIO~\cite{arcuri2018test} algorithm used in \evo applies few of these techniques.
Still, without a fitness function that can provide gradient to the search, such search would degenerate in a so called random walk on ``fitness plateau'' once a local optimum is reached, preventing further improvement.
For all these reasons, the comparisons of 2 algorithms (or algorithm variants) might give very different results based on the used search budget, especially if using different fitness functions.
A more sophisticated and expensive fitness function could give worse results for ``low'' time budgets, and better for ``higher'' budgets.

Figure~\ref{fig:ocvn} shows the results of the second set of experiments.
Those focus only on the API \emph{ocvn}, with a search budget of 10 hours.
For low search budgets, the \emph{All} configuration gives worse results.
With increasing budget, both configurations \emph{Base} and \emph{All} improve in performance, although \emph{Base} does plateau to lower values.
For line coverage, \emph{All} takes over after 1 hour.
For number of detected faults, it takes over after 9 hours.

So, based on these results, which of the 2 configurations is ``better''?
In a software engineering context,
it all depends on how common those time settings are in practice among practitioners.
Unfortunately, we do not have such information.
Most of the work done in the literature on automated testing of REST APIs has not considered its use among practitioners~\cite{golmohammadi2023testing}.
This is possibly because there is currently no popular REST API fuzzer widely used in industry.
Even when considering mature fuzzers in other popular domains, such as data parsers, there is no much research work aimed at studying how practitioners use those fuzzers~\cite{nourry2023human}.

It can be speculated that, considering the life-time of the development of a SUT which can be in years, fuzzing sessions could be relatively \emph{long}.
This is particularly the case if the fuzzing can be integrated in remote CI servers (e.g., like done in~\cite{campos2014continuous} for unit test generation), especially if the outcome of a previous fuzzing session can be reused for following sessions (e.g., using different kinds of test seeding strategies~\cite{rojas2016seeding}), even if parts of the tested code have been modified.
Easy-to-use integration of fuzzers into CI servers is a common request among users of fuzzers~\cite{nourry2023human}.
In those scenarios, more expensive techniques could pay off better in the end.
Still, hybrid approaches could be designed: e.g., simple, cheap techniques at the ``beginning'' (e.g., first few sessions on CI), followed by more expensive techniques afterwards.

Another point to consider is that, as we got better results with 10 hour budget for \emph{ocvn}, this might (or might not) happen as well for the other SUTs in our case study.
Without empirical experiments, this is not possible to tell.
Furthermore, even if there is no difference at 10 hours, perhaps there could be a difference at 24 hours, or more.
Testing all different kinds of large test budgets is unfortunately not a viable option for academic experimentation, e.g., our experiments already took more than 45 days if run in sequence.

\begin{result}
{\bf RQ2}: More sophisticated, expensive techniques can require to use longer time budgets before they can pay off.
\end{result}

\section{Discussion}
\label{sec:discussion}

Our analyses show that our novel techniques presented in this paper can improve performance significantly, in some cases.
There are several open-issues in fuzzing REST APIs~\cite{zhang2023open}, including for example how to deal with MongoDB databases and interactions with external services.
In this paper we addressed some of them, including underspecified schemas and some flag problems in existing JDK functions.
Where those problems occur, improvement can be significant.

A clear example of this is the handling of SQL schemas (Section~\ref{sec:sql}).
Based on the data in Table~\ref{tab:six}, we can see it has improvements in 2 APIs, a moderate $+2.6$\% average line coverage for \emph{catwatch}, and a more substantial $+9.6$\% coverage for \emph{cwa-verification}.
How common is such issue in practice?
Having such issue showing in 2 out of 15 APIs would mean 13\% of the employed case study.
Even if improvements can be obtained on only 13\% of APIs, there are possibly millions of APIs developed in enterprises around the world.
And so, there is practical value in these new presented techniques.
Still, EMB is not (and nor can it be) a statistically valid representative~\cite{FrA12b} of APIs built in industry, so we have no way to state how common this problem actually is in industry.
It might be less common than 13\%, or even more common.

Cherry-picking for experimentation only APIs in which improvements are obtained (in our case, 6 specific APIs) would be scientifically invalid, and ethically questionable.
To reduce bias, given a corpus like EMB, it is important to use all of it.
Even if a technique dealing with SQL databases would have for sure no benefits on databaseless APIs such as \emph{rest-ncs} and \emph{rest-scs}, it is still important to study if such a technique has no negative side-effects.
If some important API features are missing in the corpus, the corpus can be extended, as we do each year~\cite{icst2023emb,EMB}.
This also helps reducing risks of designing techniques that overfit for a specific corpus of SUTs, and so would likely not generalize to other SUTs.

Running fuzzing sessions for longer (e.g., 10 hours \emph{vs}.~1 hour) could lead to better results.
However, if the search is stuck in a local optimum with a large fitness plateau, even significantly longer time budgets would be of little help, as the search would simply degenerate into a random walk.
New techniques would be required to improve the fitness function.

\begin{figure*}[t!]
	\centering
\begin{lstlisting}[language=java]
public static boolean isValid(final String hexString) {
   if (hexString == null) {
        throw new IllegalArgumentException();
   }

   int len = hexString.length();
   if (len != 24) {  ©\label{l24}©
        return false;
   }

   for (int i = 0; i < len; i++) {
        char c = hexString.charAt(i);
        if (c >= '0' && c <= '9') {
             continue;
        }
        if (c >= 'a' && c <= 'f') {
             continue;
        }
        if (c >= 'A' && c <= 'F') {
             continue;
        }

        return false;
   }

   return true; ©\label{ltrue}©
}
\end{lstlisting}
\caption{
\label{fig:isValid}
Code of the function \texttt{isValid()} in the class \texttt{org.bson.types.ObjectId} from the library \texttt{org.mongodb:bson:3.4.2}.
Licensed under the Apache License, Version 2.0.
}
\end{figure*}

Let us consider the case of \emph{ocvn}, for which we ran a second set of experiments for 10 hour budget.
Still, achieved average line coverage does not go much over $37.1$\%.
How come?
Many of the inputs in the API calls refer to \emph{ObjectID}s for the database MongoDB.
Those input strings are marked with javax annotation constraints such as
\texttt{@EachPattern(regexp = "\^{}[a-zA-Z0-9]*\$")}.
In laymen terms, such regular expression allows for strings containing only basic letters and digits, with any size (including empty).
This is an easy to sample regex that, even without our novel techniques presented in this paper, there is no issue for \evo to sample valid values such as \texttt{1nZQfS5q}.
However, that constraint is wrong.
When generating test cases, we find many crashes (i.e., 500 status code) with error messages such as
``\texttt{invalid hexadecimal representation of an ObjectId: [1nZQfS5q]}''.
In the source code of \emph{ocvn}, such IDs are instantiated with instructions like ``\texttt{new ObjectId(s)}'', where \texttt{s} is a field in the incoming HTTP calls to the API.
The class \texttt{org.bson.types.ObjectId} from the library \texttt{org.mongodb:bson:3.4.2} does a check on the input string, as shown in Figure~\ref{fig:isValid}.
A missing constraint from the regular expression is that the input string must be of the exact length 24 (see Line~\ref{l24}).
For example, the faulty regular expression could be fixed by writing it as
\texttt{\^{}[a-zA-Z0-9]\string{24\string}\$}.

Still, even with a faulty regular expression, a white-box fuzzer should  be able to maximize code coverage.
In the case of the \texttt{if} statement with constraint \texttt{len != 24} on Line~\ref{l24}, already something basic as standard branch distance on numeric values~\cite{Kor90} should be enough to reward the search to sample strings of lengths that get closer to the value 24.
This does not happen, unfortunately, as the class \texttt{ObjectId} is not part of the business logic of the SUT, and so it is not instrumented for branch distance computations.
Indiscriminately instrumenting third-party libraries as well could have huge computational costs, as those could be millions of lines of code, even for small APIs~\cite{icst2023emb}.
Smart strategies would need to be designed (e.g., to avoid instrumenting classes that have no impact on the execution flow of the business logic of the API).
Furthermore, we would not want to cover the \texttt{true} output (i.e., Line~\ref{ltrue}) just once, but in all places from which  ``\texttt{new ObjectId(s)}'' is called from.
This most likely would require inter-method call heuristics, e.g., by adapting techniques presented for unit testing such as~\cite{vogl2021encoding,lin2020recovering}.

An alternative approach could be to provide method replacements for \texttt{ObjectId}, like done for JDK classes in Section~\ref{sec:new_replacements}.
On the one hand, providing adhoc transformations for each single method for the vast amount of third-party libraries out there in the world would be an infeasible task.
A more general solution would be preferred (e.g., as experimented for unit testing in~\cite{vogl2021encoding,lin2020recovering}).
On the other hand, for widely used libraries, and widely used classes/functions in them, it could make sense to provide adhoc solutions.
An adhoc solution would be more efficient than a generic one.
For example, a method replacement for \texttt{ObjectId} could check for tainted values, and treat them like they were matched with the regular expression
\texttt{\^{}[a-zA-Z0-9]\string{24\string}\$}.
This could significantly boost the search.

\section{Threats To Validity}
\label{sec:threats}

To address threats to internal validity,
our code implementation has been carefully tested, with several unit and end-to-end tests~\cite{arcuri2023building}.
Furthermore, \evo is open-source, with all the new releases automatically stored on Zenodo for long term storage, like for example version 1.6.1~\cite{zenodo161evomaster}.
Anyone can review its source code.
Furthermore, most of our case study is from the open-source corpus EMB~\cite{icst2023emb,EMB}, which is as well stored on Zenodo.
However, we are not allowed to share the API of our industrial partner.

To handle the randomness of the employed algorithms, all experiments were repeated 10 times, and analyzed with the appropriate statistical tests, following common guidelines in the literature~\cite{Hitchhiker14}.

In this paper, we have not compared our novel white-box techniques with any other white-box fuzzer, as none exists for RESTful APIs besides \evo~\cite{golmohammadi2023testing}.
Comparing with existing black-box techniques would add only little to the discussion, as there are already few studies that show that \evo even in its black-box mode gives the best results, and white-box testing, when applicable, significantly improves over black-box testing.
The reader interested in tool comparisons is referred to our previous work in~\cite{zhang2023open}, and the independent study carried out by Kim \emph{et~al.}~\cite{Kim2022Rest}.

Regarding threats to external validity, we used  all the REST APIs in the EMB corpus, plus one API from one of our industrial partners.
This provides a wide range of different APIs in terms of complexity and type.
Still, as for most empirical studies, we cannot guarantee that our results would generalize to other APIs as well.

Our empirical study focused on REST APIs.
However, some of our techniques could be applicable and successful in other domains as well.
For example, the handling of underspecified SQL schemas could be applicable to other types of enterprise systems that use databases as well, like for example GraphQL and RPC-based APIs.
Method replacements to address the flag problem could be used also in other white-box testing contexts, such as for example unit test generation.
However, without proper empirical studies, whether our novel techniques would be successful in those cases as well is not something that can be given for granted.

\section{Conclusions}
\label{sec:conclusions}

There is a large body of research in software test generation, with successful applications in many different contexts.
When it comes to test enterprise systems such as REST APIs, besides our own work with \evo~\cite{arcuri2021evomaster}, all work in the literature has been aimed at black-box testing~\cite{golmohammadi2023testing}.
White-box testing can significantly improve performance on achieved code coverage and fault detection, especially for REST APIs~\cite{arcuri2020blackbox,zhang2023open}.
Considering the widespread use of web services in industry, more research on this topic is warranted.

In this paper we provided novel search-based heuristics to push forward the start-of-the-art in this software testing domain.
Experiments on 14 open-source and 1 industrial APIs show the effectiveness of our novel techniques.
For example, on a service from the German Covid's app backend (i.e., \emph{cwa-verification}), our novel handling of underspecified SQL schemas increased average line coverage by $9.6$\%.

Still, considering an average line coverage of $56.1$\% with a 1-hour search budget, there are several research challenges that need to be overcome in the future.
Several of these issues have been already identified in the literature~\cite{zhang2023open}, including how to deal with interactions with external services and databases such as MongoDB.
As all of our work is open-source~\cite{zenodo161evomaster}, it can be used to bootstrap new research effort in this domain.

What presented in this paper is active by default since \evo version 1.6.1.
At the time of this writing, based on download statistics and direct contact with large companies such as Meituan~\cite{meituanArxiv2022}, hundreds of engineers in industry are already benefiting from the scientific research presented in this paper.
\evo is open-source on GitHub~\cite{EvoMaster} with releases automatically stored on Zenodo~\cite{zenodo161evomaster}.
To enable the replicability of our scientific studies, the code repository of \evo also contains the scripts used to run experiments, and documentation on how to use them.

\section*{Acknowledgments}
This work is funded by the European Research Council (ERC) under the European Union’s Horizon 2020 research and innovation programme (EAST project, grant agreement No. 864972), and partially funded by UBACYT-2020 20020190100233BA, PICT-2019-01793.


\bibliographystyle{ACM-Reference-Format} 

%



\section*{APPENDIX}

In this appendix, we provide extra tables with more details on the comparisons of our novel techniques.

\begin{table*}
	\centering
	\caption{
	    Same kind of analysis done in Table~\ref{tab:all}, but for the configuration \emph{TAOS}.
	}
	\label{tab:taos}
	\resizebox{1.\linewidth}{!}{
		\begin{tabular}{ l rrrr rrrr rrr}\\ 
\toprule 
SUT & \multicolumn{4}{c}{Line Coverage \%} & \multicolumn{4}{c}{\# Detected Faults} & \multicolumn{3}{c}{\# HTTP Calls} \\ 
    & Base & TAOS  & $\hat{A}_{12}$ & p-value  & Base & TAOS & $\hat{A}_{12}$ & p-value & Base & TAOS & Difference \\ 
\midrule 
\emph{catwatch} & 42.5 & 41.9 & 0.40 & 0.479 & 19.4 & 18.7 & 0.33 & 0.241 & 3120 & 2669 & +14.43\\
\emph{cwa-verification} & 47.4 & 47.5 & 0.70 & 0.177 & 7.4 & 7.4 & 0.50 & 1.000 & 151478 & 150240 & +0.82\\
\emph{features-service} & 81.5 & 81.5 & 0.57 & 0.604 & 33.2 & 33.9 & 0.49 & 0.967 & 182534 & 182441 & +0.05\\
\emph{genome-nexus} & 36.7 & 37.3 & 0.71 & 0.130 & 20.0 & 20.6 & 0.60 & 0.460 & 59304 & 52567 & +11.36\\
\emph{gestaohospital-rest} & 39.6 & 39.5 & 0.35 & 0.204 & 22.0 & 22.0 & 0.50 & 1.000 & 240280 & 239161 & +0.47\\
\emph{ind0} & 12.4 & 24.1 & {\bf 0.99} & $< 0.001$ & 44.1 & 54.7 & {\bf 1.00} & $< 0.001$ & 232335 & 184305 & +20.67\\
\emph{languagetool} & 41.7 & 41.0 & 0.42 & 0.579 & 8.3 & 5.4 & {\bf 0.23} & 0.035 & 13713 & 11052 & +19.41\\
\emph{market} & 48.6 & 47.6 & 0.29 & 0.118 & 20.0 & 19.6 & 0.40 & 0.461 & 27731 & 27520 & +0.76\\
\emph{ocvn-rest} & 37.1 & 37.1 & 0.49 & 0.968 & 553.8 & 515.3 & {\bf 0.10} & 0.003 & 136751 & 143464 & -4.91\\
\emph{proxyprint} & 53.2 & 52.9 & 0.53 & 0.847 & 83.6 & 86.9 & 0.65 & 0.310 & 40095 & 38109 & +4.95\\
\emph{rest-ncs} & 93.0 & 93.0 & 0.50 & 1.000 & 6.0 & 6.0 & 0.50 & 1.000 & 275741 & 276332 & -0.21\\
\emph{rest-news} & 66.9 & 67.4 & 0.73 & 0.063 & 8.0 & 8.0 & 0.50 & 1.000 & 300261 & 297393 & +0.96\\
\emph{rest-scs} & 85.7 & 85.7 & 0.51 & 0.968 & 12.0 & 11.9 & 0.45 & 0.368 & 283604 & 275094 & +3.00\\
\emph{restcountries} & 77.0 & 77.1 & {\bf 0.75} & 0.014 & 2.0 & 2.0 & 0.50 & 1.000 & 233799 & 233477 & +0.14\\
\emph{scout-api} & 52.9 & 52.7 & 0.41 & 0.496 & 89.5 & 89.3 & 0.49 & 0.970 & 157931 & 157212 & +0.46\\
\midrule 
Average  & 54.4 & 55.1 & 0.56 &  & 62.0 & 60.1 & 0.48 &  & 155912 & 151402 & +4.82\\
Median  & 48.6 & 47.6 & 0.51 &  & 20.0 & 19.6 & 0.50 &  & 157931 & 157212 & +0.82\\
\bottomrule 
\end{tabular} 

	}
\end{table*}

\begin{table*}
	\centering
	\caption{
	    Same kind of analysis done in Table~\ref{tab:all}, but for the configuration \emph{TT}.
	}
	\label{tab:tt}
	\resizebox{1.\linewidth}{!}{
		\begin{tabular}{ l rrrr rrrr rrr}\\ 
\toprule 
SUT & \multicolumn{4}{c}{Line Coverage \%} & \multicolumn{4}{c}{\# Detected Faults} & \multicolumn{3}{c}{\# HTTP Calls} \\ 
    & Base & TT  & $\hat{A}_{12}$ & p-value  & Base & TT & $\hat{A}_{12}$ & p-value & Base & TT & Difference \\ 
\midrule 
\emph{catwatch} & 42.5 & 45.7 & {\bf 0.89} & 0.005 & 19.4 & 25.3 & {\bf 1.00} & $< 0.001$ & 3120 & 12948 & -315.04\\
\emph{cwa-verification} & 47.4 & 47.6 & 0.85 & 0.098 & 7.4 & 8.2 & 0.85 & 0.075 & 151478 & 134539 & +11.18\\
\emph{features-service} & 81.5 & 81.5 & 0.46 & 0.795 & 33.2 & 34.8 & 0.68 & 0.188 & 182534 & 198295 & -8.63\\
\emph{genome-nexus} & 36.7 & 36.9 & 0.57 & 0.650 & 20.0 & 20.9 & 0.66 & 0.241 & 59304 & 44243 & +25.40\\
\emph{gestaohospital-rest} & 39.6 & 39.6 & 0.40 & 0.398 & 22.0 & 22.0 & 0.50 & 1.000 & 240280 & 228894 & +4.74\\
\emph{ind0} & 12.4 & 15.6 & 0.74 & 0.094 & 44.1 & 50.6 & {\bf 0.99} & $< 0.001$ & 232335 & 204282 & +12.07\\
\emph{languagetool} & 41.7 & 41.1 & 0.32 & 0.190 & 8.3 & 7.4 & 0.42 & 0.540 & 13713 & 14234 & -3.79\\
\emph{market} & 48.6 & 47.8 & 0.47 & 0.855 & 20.0 & 19.3 & 0.33 & 0.207 & 27731 & 28260 & -1.91\\
\emph{ocvn-rest} & 37.1 & 37.1 & 0.59 & 0.499 & 553.8 & 539.9 & {\bf 0.05} & 0.001 & 136751 & 118203 & +13.56\\
\emph{proxyprint} & 53.2 & 51.5 & 0.34 & 0.336 & 83.6 & 82.6 & 0.41 & 0.600 & 40095 & 35266 & +12.04\\
\emph{rest-ncs} & 93.0 & 93.0 & 0.50 & 1.000 & 6.0 & 6.0 & 0.50 & 1.000 & 275741 & 272240 & +1.27\\
\emph{rest-news} & 66.9 & 67.6 & {\bf 0.78} & 0.025 & 8.0 & 7.8 & 0.40 & 0.167 & 300261 & 299284 & +0.33\\
\emph{rest-scs} & 85.7 & 86.0 & 0.63 & 0.308 & 12.0 & 12.0 & 0.50 & 1.000 & 283604 & 268197 & +5.43\\
\emph{restcountries} & 77.0 & 77.1 & 0.65 & 0.210 & 2.0 & 2.0 & 0.50 & 1.000 & 233799 & 221756 & +5.15\\
\emph{scout-api} & 52.9 & 53.0 & 0.38 & 0.404 & 89.5 & 90.6 & 0.55 & 0.733 & 157931 & 154334 & +2.28\\
\midrule 
Average  & 54.4 & 54.7 & 0.57 &  & 62.0 & 62.0 & 0.56 &  & 155912 & 148998 & -15.73\\
Median  & 48.6 & 47.8 & 0.57 &  & 20.0 & 20.9 & 0.50 &  & 157931 & 154334 & +4.74\\
\bottomrule 
\end{tabular} 

	}
\end{table*}

\begin{table*}
	\centering
	\caption{
	    Same kind of analysis done in Table~\ref{tab:all}, but for the configuration \emph{TT+OpenAPI}.
	}
	\label{tab:openapi}
	\resizebox{1.\linewidth}{!}{
		\begin{tabular}{ l rrrr rrrr rrr}\\ 
\toprule 
SUT & \multicolumn{4}{c}{Line Coverage \%} & \multicolumn{4}{c}{\# Detected Faults} & \multicolumn{3}{c}{\# HTTP Calls} \\ 
    & Base & TT+OpenAPI  & $\hat{A}_{12}$ & p-value  & Base & TT+OpenAPI & $\hat{A}_{12}$ & p-value & Base & TT+OpenAPI & Difference \\ 
\midrule 
\emph{catwatch} & 42.5 & 47.9 & {\bf 0.96} & $< 0.001$ & 19.4 & 26.2 & {\bf 1.00} & $< 0.001$ & 3120 & 14493 & -364.56\\
\emph{cwa-verification} & 47.4 & 47.6 & {\bf 0.94} & 0.019 & 7.4 & 8.0 & 0.80 & 0.067 & 151478 & 127285 & +15.97\\
\emph{features-service} & 81.5 & 81.4 & 0.44 & 0.636 & 33.2 & 33.2 & 0.42 & 0.582 & 182534 & 190030 & -4.11\\
\emph{genome-nexus} & 36.7 & 36.4 & 0.43 & 0.623 & 20.0 & 20.5 & 0.56 & 0.696 & 59304 & 43972 & +25.85\\
\emph{gestaohospital-rest} & 39.6 & 39.6 & 0.45 & 0.681 & 22.0 & 22.0 & 0.50 & 1.000 & 240280 & 230234 & +4.18\\
\emph{ind0} & 12.4 & 13.5 & 0.48 & 0.930 & 44.1 & 49.1 & {\bf 0.84} & 0.016 & 232335 & 208888 & +10.09\\
\emph{languagetool} & 41.7 & 41.9 & 0.49 & 0.971 & 8.3 & 11.2 & 0.72 & 0.101 & 13713 & 28754 & -109.68\\
\emph{market} & 48.6 & 47.2 & 0.31 & 0.186 & 20.0 & 19.9 & 0.47 & 0.874 & 27731 & 27516 & +0.78\\
\emph{ocvn-rest} & 37.1 & 37.1 & {\bf 0.95} & $< 0.001$ & 553.8 & 547.0 & 0.31 & 0.177 & 136751 & 115378 & +15.63\\
\emph{proxyprint} & 53.2 & 54.1 & 0.62 & 0.491 & 83.6 & 86.2 & 0.57 & 0.696 & 40095 & 40491 & -0.99\\
\emph{rest-ncs} & 93.0 & 93.0 & 0.50 & 1.000 & 6.0 & 6.0 & 0.50 & 1.000 & 275741 & 258372 & +6.30\\
\emph{rest-news} & 66.9 & 67.3 & 0.66 & 0.210 & 8.0 & 7.8 & 0.40 & 0.167 & 300261 & 289447 & +3.60\\
\emph{rest-scs} & 85.7 & 86.2 & 0.70 & 0.119 & 12.0 & 12.0 & 0.50 & 1.000 & 283604 & 269067 & +5.13\\
\emph{restcountries} & 77.0 & 77.0 & 0.52 & 0.899 & 2.0 & 2.0 & 0.50 & 1.000 & 233799 & 223623 & +4.35\\
\emph{scout-api} & 52.9 & 53.2 & 0.43 & 0.623 & 89.5 & 89.4 & 0.46 & 0.791 & 157931 & 139254 & +11.83\\
\midrule 
Average  & 54.4 & 54.9 & 0.59 &  & 62.0 & 62.7 & 0.57 &  & 155912 & 147120 & -25.04\\
Median  & 48.6 & 47.9 & 0.50 &  & 20.0 & 20.5 & 0.50 &  & 157931 & 139254 & +4.35\\
\bottomrule 
\end{tabular} 

	}
\end{table*}

\begin{table*}
	\centering
	\caption{
	    Same kind of analysis done in Table~\ref{tab:all}, but for the configuration \emph{JPA}.
	}
	\label{tab:jpa}
	\resizebox{1.\linewidth}{!}{
		\begin{tabular}{ l rrrr rrrr rrr}\\ 
\toprule 
SUT & \multicolumn{4}{c}{Line Coverage \%} & \multicolumn{4}{c}{\# Detected Faults} & \multicolumn{3}{c}{\# HTTP Calls} \\ 
    & Base & JPA  & $\hat{A}_{12}$ & p-value  & Base & JPA & $\hat{A}_{12}$ & p-value & Base & JPA & Difference \\ 
\midrule 
\emph{catwatch} & 42.5 & 45.1 & {\bf 0.84} & 0.014 & 19.4 & 19.5 & 0.48 & 0.932 & 3120 & 3574 & -14.55\\
\emph{cwa-verification} & 47.4 & 57.0 & {\bf 1.00} & 0.017 & 7.4 & 10.2 & {\bf 0.95} & 0.031 & 151478 & 137342 & +9.33\\
\emph{features-service} & 81.5 & 81.8 & 0.62 & 0.375 & 33.2 & 35.5 & {\bf 0.78} & 0.033 & 182534 & 179841 & +1.47\\
\emph{genome-nexus} & 36.7 & 36.3 & 0.44 & 0.677 & 20.0 & 20.1 & 0.51 & 1.000 & 59304 & 54399 & +8.27\\
\emph{gestaohospital-rest} & 39.6 & 39.4 & {\bf 0.25} & 0.032 & 22.0 & 22.0 & 0.50 & 1.000 & 240280 & 234277 & +2.50\\
\emph{ind0} & 12.4 & 13.9 & 0.59 & 0.513 & 44.1 & 45.0 & 0.50 & 1.000 & 232335 & 219452 & +5.55\\
\emph{languagetool} & 41.7 & 38.3 & 0.39 & 0.436 & 8.3 & 7.9 & 0.38 & 0.356 & 13713 & 11321 & +17.45\\
\emph{market} & 48.6 & 47.2 & 0.26 & 0.080 & 20.0 & 19.4 & 0.37 & 0.347 & 27731 & 28192 & -1.66\\
\emph{ocvn-rest} & 37.1 & 37.1 & 0.54 & 0.746 & 553.8 & 550.7 & 0.32 & 0.197 & 136751 & 138513 & -1.29\\
\emph{proxyprint} & 53.2 & 54.0 & 0.65 & 0.321 & 83.6 & 82.1 & 0.41 & 0.561 & 40095 & 40918 & -2.05\\
\emph{rest-ncs} & 93.0 & 93.0 & 0.50 & 1.000 & 6.0 & 6.0 & 0.50 & 1.000 & 275741 & 271886 & +1.40\\
\emph{rest-news} & 66.9 & 66.8 & 0.47 & 0.824 & 8.0 & 7.6 & {\bf 0.30} & 0.034 & 300261 & 290365 & +3.30\\
\emph{rest-scs} & 85.7 & 86.3 & 0.74 & 0.064 & 12.0 & 12.0 & 0.50 & 1.000 & 283604 & 277050 & +2.31\\
\emph{restcountries} & 77.0 & 77.0 & 0.60 & 0.425 & 2.0 & 2.0 & 0.50 & 1.000 & 233799 & 230260 & +1.51\\
\emph{scout-api} & 52.9 & 54.4 & 0.62 & 0.384 & 89.5 & 90.1 & 0.48 & 0.910 & 157931 & 148864 & +5.74\\
\midrule 
Average  & 54.4 & 55.2 & 0.57 &  & 62.0 & 62.0 & 0.50 &  & 155912 & 151083 & +2.62\\
Median  & 48.6 & 54.0 & 0.59 &  & 20.0 & 19.5 & 0.50 &  & 157931 & 148864 & +2.31\\
\bottomrule 
\end{tabular} 

	}
\end{table*}

\end{document}